\documentclass[a4paper,11pt]{article}
\pdfoutput=1 % if your are submitting a pdflatex (i.e. if you have
\usepackage{jheppub} % for details on the use of the package, please
\usepackage[utf8]{inputenc}
\usepackage[T1]{fontenc} % if needed
\usepackage{booktabs} % if needed
\usepackage{placeins}
\usepackage{float}
\usepackage{changes,todonotes}
\usepackage{tikz}
\usetikzlibrary{arrows}
\usetikzlibrary{shapes,arrows}
\usetikzlibrary{positioning}
\usetikzlibrary{fit}
\usetikzlibrary{shadows}

\usepackage{amsmath,amssymb,amsfonts}	% DON'T use cite, screws with bibtex + hyperref
\usepackage{color}
\usepackage{amsmath}
\usepackage{amsfonts}
\usepackage{amssymb}
\usepackage{graphicx}
\usepackage{slashed}            % for slashed characters in math mode
\usepackage{bbm}                % for \mathbbm{1} (unit matrix)
\usepackage{xspace}				% For spacing after commands
\usepackage{tikz}
\usetikzlibrary{arrows,shapes}
\usetikzlibrary{trees}
\usetikzlibrary{matrix,arrows} 				% For commutative diagram
\usetikzlibrary{positioning}				% For "above of=" commands
\usetikzlibrary{calc,through}				% For coordinates
\usetikzlibrary{decorations.pathreplacing}  % For curly braces
\usepackage{pgffor}							% For repeating patterns
\usetikzlibrary{decorations.pathmorphing}	% For Feynman Diagrams
\usetikzlibrary{decorations.markings}
\tikzset{
    vector/.style={decorate, decoration={snake}, draw},
	provector/.style={decorate, decoration={snake,amplitude=2.5pt}, draw},
	antivector/.style={decorate, decoration={snake,amplitude=-2.5pt}, draw},
    fermion/.style={draw=black, postaction={decorate},
        decoration={markings,mark=at position .55 with {\arrow[draw=black]{>}}}},   
    fermionbar/.style={draw=black, postaction={decorate},
        decoration={markings,mark=at position .55 with {\arrow[draw=black]{<}}}},
    fermionnoarrow/.style={draw=black},
    gluon/.style={decorate, draw=black,
        decoration={coil,aspect=0.5,amplitude=4pt, segment length=4pt}},
    scalar/.style={dashed,draw=black, postaction={decorate},
        decoration={markings,mark=at position .55 with {\arrow[draw=black]{>}}}},
    scalarbar/.style={dashed,draw=black, postaction={decorate},
        decoration={markings,mark=at position .55 with {\arrow[draw=black]{<}}}},
    scalarnoarrow/.style={dashed,draw=black},
    electron/.style={draw=black, postaction={decorate},
        decoration={markings,mark=at position .55 with {\arrow[draw=black]{>}}}},
	bigvector/.style={decorate, decoration={snake,amplitude=4pt}, draw},
	ghost/.style={dotted,draw=black, postaction={decorate},
        decoration={markings,mark=at position .55 with {\arrow[draw=black]{>}}}},
	ghostbar/.style={dotted,draw=black, postaction={decorate},
        decoration={markings,mark=at position .55 with {\arrow[draw=black]{<}}}},
    feynarrow/.style={draw=black, postaction={decorate},
        decoration={markings,mark=at position 1.0 with {\arrow[draw=black]{>}}}},
         gluonloop/.style={decorate, draw=black,
        decoration={coil,aspect=0.6,amplitude=3pt, segment length=4pt}},   
}
\usepackage[]{units}
\usepackage{tensor}
\usepackage{tikz-3dplot}

\usepackage{pdflscape}

\title{Resonance-aware subtraction in the dipole method}
\preprint{SLAC-PUB-17299\\
  \hphantom{*}\hfill FERMILAB-PUB-18-764-T\\
  \hphantom{*}\hfill MCNET-18-15}

\author[a,b]{Stefan H\"oche,}
\author[c]{Sebastian Liebschner,}
\author[c]{and Frank Siegert}

\affiliation[a]{SLAC National Accelerator Laboratory, Menlo Park, CA, 94025, USA}
\affiliation[b]{Fermi National Accelerator Laboratory, Batavia, IL, 60510-0500, USA}
\affiliation[c]{Institut f\"ur Kern- und Teilchenphysik, TU Dresden,
 01069 Dresden, Germany}

\emailAdd{shoeche@slac.stanford.edu}
\emailAdd{sebastian.liebschner@tu-dresden.de}
\emailAdd{frank.siegert@cern.ch}
\abstract{We present a technique for infrared subtraction in next-to-leading
  order QCD calculations that preserves the virtuality of resonant propagators.
  The approach is based on the pseudo-dipole subtraction method proposed
  by Catani and Seymour in the context of identified particle production.
  As the first applications, we compute the $e^+e^- \to W^+W^-b\bar{b}$ and
  $pp \to W^+W^-j_bj_b$ cross-section, which are both dominated by top-quark
  pair production above the threshold.
  We compare the efficiency of our approach with a calculation performed
  using the standard dipole subtraction technique.}
\begin{document} 
\maketitle
\flushbottom

\section{Introduction}
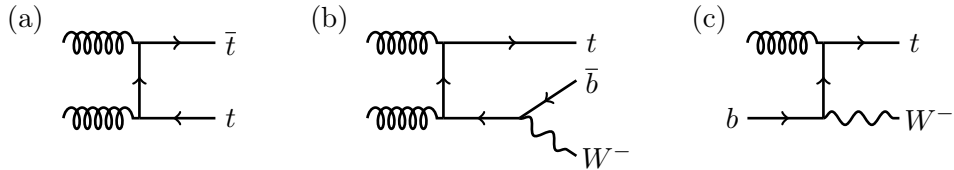
\begin{figure}[t]
\begin{center} 
\begin{tikzpicture}[line width=1.0 pt, scale=1.0]
\begin{scope}[shift={(9,0)}]
    \node at (-1.5,0.8) {(c)};
	\draw[gluon] (-1,0.5)--(0,.5);
	\draw[fermion] (-1,-0.5)--(0,-.5);
	\draw[fermion] (0,-0.5)--(0,.5);
	\draw[fermion] (0,0.5)--(1,.5);
	\draw[vector] (0,-0.5)--(1,-.5);
	\node at (-1.2,-0.5) {$b$};
	\node at (1.2,0.5) {$t$};
	\node at (1.4,-0.5) {$W^-$};
\end{scope}	
\begin{scope}[shift={(4,0)}]
	\node at (-1.5,0.8) {(b)};
	\draw[gluon] (-1,0.5)--(0,.5);
	\draw[gluon] (-1,-0.5)--(0,-.5);
	\draw[fermion] (0,-0.5)--(0,.5);
	\draw[fermion] (0,0.5)--(1.75,.5);
	\draw[fermionbar] (0,-0.5)--(1,-0.5);
	\draw[fermionbar] (1,-0.5)--(1.75,0.0);
	\draw[vector] (1,-0.5)--(1.75,-1.0);
	\node at (1.95,0.5) {$t$};
	\node at (1.95,0) {$\overline{b}$};
	\node at (2.15,-1.0) {$W^-$};
\end{scope}	
	\node at (-1.5,0.8) {(a)};
	\draw[gluon] (-1,0.5)--(0,.5);
	\draw[gluon] (-1,-0.5)--(0,-.5);
	\draw[fermion] (0,-0.5)--(0,.5);
	\draw[fermion] (0,0.5)--(1,.5);
	\draw[fermionbar] (0,-0.5)--(1,-.5);
	\node at (1.2,0.5) {$\overline{t}$};
	\node at (1.2,-0.5) {$t$};
\end{tikzpicture}
\end{center}
\caption{Representative diagrams for (a) single-top-production, (b) single-top-production +jet and (c) $t\overline{t}$-production}\label{fig:overlap}
\end{figure}

The production and decay of heavy resonances like the top quark is of
greatest interest in particle-physics phenomenology~\cite{Agashe:2013hma}.
It presents a window into new physics, which is commonly believed to emerge
in the form of new interactions at high energy. Precision measurements of
Standard Model parameters at current collider energies may reveal parts
of this structure if they can be made at high precision.
The necessary theoretical predictions for top-quark pair production have been
computed at next-to-leading order (NLO)~\cite{Nason:1989zy,Beenakker:1990maa,
  Mangano:1991jk,Frixione:1995fj} and next-to-next-to leading order (NNLO)%
~\cite{Baernreuther:2012ws,Czakon:2012pz,Czakon:2013goa} QCD perturbation theory,
and combined with NLO electroweak results~\cite{Czakon:2017wor}.
In these calculations, top quarks are considered to be asymptotic final states,
and finite width effects are neglected. When including top quark decays,
a problem arises that is related to the very definition of the inclusive
final state and can most easily be explained using Fig.~\ref{fig:overlap}.
Diagram~(a) represents the top-quark pair production process in leading order
perturbation theory. Diagram~(b) can be obtained from diagram~(a) by including
the decay of one of top quarks. It may also be considered as a real radiative
correction to the single-top quark production process represented by diagram~(c).
Quite obviously, diagram~(b) is resonant when $|( p_W + p_b )^2 - m_t^2|\lessapprox \Gamma_t^2$.
In this region of phase space the NLO calculation of $pp\to Wt$ therefore overlaps
with the calculation of $pp\to t[\bar{t}\to Wb]$ and spoils the definition of
a NLO cross section for $pp\to Wt$. This problem has traditionally been addressed by techniques
such as diagram removal or diagram subtraction~\cite{GoncalvesNetto:2012yt}.
However, both methods introduce theoretical uncertainties
and violate gauge invariance. The natural approach is instead to not view
single-top and top-pair production as two separable channels and to
consider only the fully decayed final state~\cite{Denner:2010jp,
  Bevilacqua:2010qb,Denner:2012yc}.
In addition to being theoretically robust, this technique matches the reality
of performing an experimental measurement, where the existence of the top quark
as an intermediate state is inferred from the decay products.

Higher-order radiative corrections to both the production and the decay
of top quarks can also be simulated numerically in computer programs
called event generators, which allow to map experimental signatures associated with
top-quark production to the parameters of the theory. It is the factorized approach
of these simulations that presents a problem when the precision target of the experimental
measurement lies below the resonance width, because the narrow width approximation
can no longer be applied~\cite{Heinrich:2013qaa}. Again, the natural solution
is to perform the computation for the complete final state and match the NLO fixed-order result
to the parton shower~\cite{Frixione:2002ik,Frixione:2007nw}.
Since the process is an interplay of continuum contributions and resonant top-quark
production as in Fig.~\ref{fig:overlap}~(c), it could in principle be treated
in the narrow-width approximation if $|( p_W + p_b )^2 - m_t^2|\lessapprox \Gamma_t^2$.
This mandates a special choice of kinematics mapping in the transition from Born to real-emission
final states in the matching procedure, which has been discussed in great detail
\cite{Jezo:2015aia,Jezo:2016ujg,Frederix:2016rdc,Ravasio:2018lzi,Jezo:2018yaf}
in the context of the Frixione-Kunszt-Signer subtraction method~\cite{Frixione:1995ms}.
However, no attempt has so far been made to implement a solution based on
Catani-Dittmaier-Seymour-Trocsanyi dipole subtraction~\cite{Catani:1996vz,Catani:2002hc}.
In this manuscript we therefore discuss a new technique that is based
on the identified particle methods presented in~\cite{Catani:1996vz} and apply the
procedure to the computation of top-quark pair production at a future linear
collider~\cite{Fujii:2015jha,Moortgat-Picka:2015yla} and at the Large Hadron Collider (LHC),
where measurements of singly- and doubly-resonant top-quark pair production have just been
reported~\cite{Aaboud:2018bir}.

The outline of this paper is as follows: Section~\ref{sec:pseudo_dipoles} introduces
the problem of resonances in NLO calculations, reviews the pseudo-dipole subtraction
formalism as introduced in~\cite{Catani:1996vz} and shows how it
can be applied to resonance-aware subtraction. Section~\ref{sec:ee_bwbw}
presents first applications, and Sec.~\ref{sec:summary} gives an outlook.

\section{Pseudo-dipole subtraction}
\label{sec:pseudo_dipoles}
The calculation of observables at NLO requires the computation of real and virtual corrections to the Born cross section. 
After renormalisation both these contributions are still separately infinite, although their sum is finite for
infrared safe observables. In order to calculate such observables efficiently, general next-to-leading order
infrared subtraction schemes have been devised, the most widely used being the methods by Frixione, Kunszt and Signer
(FKS)~\cite{Frixione:1995ms} and the ones by Catani and Seymour (CS)~\cite{Catani:1996vz,Catani:2002hc}.
Both methods are based on the extraction of the singular limits of the real-emission corrections, their analytic
integration and combination of the result with the virtual corrections to render both real-emission and virtual
corrections separately infrared finite.
Focusing, for simplicity, on the total cross section in a process with no initial state hadrons, we can write schematically
\begin{align}
  \sigma^{\mathrm{NLO}} = \int_{m} \left( \mathrm{d}\sigma^\mathrm{B} +\mathrm{d}\sigma^\mathrm{V} + \mathrm{d}\sigma^\mathrm{I} \right)
  + \int_{m+1} \left( \mathrm{d}\sigma^{\mathrm{R}} - \mathrm{d}\sigma^{\mathrm{S}} \right).
\end{align}
Here $\int_{m} \mathrm{d}\sigma^\mathrm{I} = \int_{m+1}  \mathrm{d}\sigma^{\mathrm{S}}$ is the subtraction term,
which is analytically integrated over the phase-space of the additional parton in the real correction and $\int_m$
indicates that the phase-space integral corresponds to $m$ final-state partons.\footnote{Our analysis focuses on NLO QCD corrections,
  though the method can be extended to include NLO electroweak corrections.} 
In the remainder of this paper we will focus on CS dipole subtraction~\cite{Catani:1996vz}.
In processes with intermediate resonances, this technique exhibits an undesired feature, which
can most easily be explained using a concrete example, say $e^+e^- \to W^+W^-b\bar{b}$.
If the center-of-mass energy is greater than the top pair threshold $\sqrt{s} \gtrsim 2m_t$,
this process is dominated by on-shell $t\bar{t}$-production and decay.
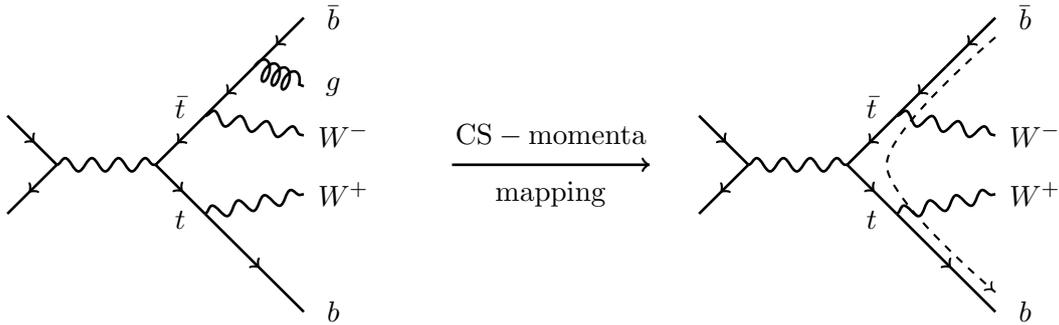
\begin{figure}[t]
\begin{center}
\begin{tikzpicture}[line width=1.0 pt, scale=1.3, line/.style={dashed,thick,draw=black, postaction={decorate},
        decoration={markings,mark=at position 0.99 with {\arrow[draw=black]{>}}}}]
	\draw[fermion] (-1,0.5)--(-0.5,0);
	\draw[fermionbar] (-1,-0.5)--(-0.5,0);	
	\draw[vector] (-0.5,0)--(0.5,0); 
	\draw[fermionbar] (0.5,0)--node[label=above:$\bar{t}$]{}(1,0.5);
	\draw[fermionbar] (1,0.5)--(1.5,1);
	\draw[fermionbar] (1.5,1)--(2,1.5);
	\draw[gluon] (1.5,1)--(2,0.8);
	\draw[vector] (1,0.5)--(2,0.3);
	\draw[fermion] (0.5,0)--node[label=below:$t$]{}(1,-0.5);
	\draw[fermion] (1,-0.5)--(2,-1.5);
	\draw[vector] (1,-0.5)--(2,-0.3);
	\node at (2.3,1.5) {$\bar{b}$};
	\node at (2.3,0.8) {$g$};
	\node at (2.4,0.3) {$W^-$};
	\node at (2.4,-0.3) {$W^+$};
	\node at (2.3,-1.5) {$b$};
	\draw[feynarrow] (3.5,0)--(5.5,0);
\node at (4.5,0.3){$\mathrm{CS-momenta}$};
\node at (4.5,-0.3){$\mathrm{mapping}$};
\begin{scope}[shift={(7,0)}]
	\draw[fermion] (-1,0.5)--(-0.5,0);
	\draw[fermionbar] (-1,-0.5)--(-0.5,0);	
	\draw[vector] (-0.5,0)--(0.5,0); 
	\draw[fermionbar] (0.5,0)--node[label=above:$\bar{t}$]{}(1,0.5);
	\draw[fermionbar] (1,0.5)--(1.5,1);
	\draw[fermionbar] (1.5,1)--(2,1.5);
	\draw[vector] (1,0.5)--(2,0.3);
	\draw[fermion] (0.5,0)--node[label=below:$t$]{}(1,-0.5);
	\draw[fermion] (1,-0.5)--(2,-1.5);
	\draw[vector] (1,-0.5)--(2,-0.3);
	\node at (2.3,1.5) {$\bar{b}$};
	\node at (2.4,0.3) {$W^-$};
	\node at (2.4,-0.3) {$W^+$};
	\node at (2.3,-1.5) {$b$};
	\draw[line] plot [smooth] coordinates { (2.0,1.3) (0.9,0.0)
	  (2.0,-1.3)};
\end{scope}
\end{tikzpicture}
\end{center}
\caption{Possible real correction configuration for $W^+W^-b\bar{b}$ production and Born configuration of associated standard CS-dipole. 
The curved arrow on the right indicates the flow of the recoil.}\label{fig:assignment_CSexample}
\end{figure}
One possible real emission correction to this process is depicted on the left-hand side of Fig.~\ref{fig:assignment_CSexample}. 
The subtraction term associated to the $\bar{b}g$ collinear sector is constructed from the Born-diagram on the right-hand side of
Fig.~\ref{fig:assignment_CSexample}, and its kinematics is obtained by mapping the on-shell final-state momenta of the real correction
to Born kinematics using the algorithm in~\cite{Catani:1996vz}. In the canonical method, the momenta of the emitter (the $\bar{b}$-quark)
and the spectator (the $b$-quark) are adjusted, while all other momenta remain the same. This procedure generates a recoil that is 
indicated by the dashed line in Fig.~\ref{fig:assignment_CSexample}. The recoil leads to the subtraction term being evaluated
at different virtualities of the intermediate top-quarks than the real-emission diagrams whose divergences it counteracts. 
As the top-quark propagator scales like $(p_t^2 - m_t^2 + im_t\Gamma_t)^{-1}$
%\begin{align*}
%\frac{1}{p_t^2 - m_t^2 + im_t\Gamma_t},
%\end{align*}
and $\Gamma_t \ll m_t$, the change in virtuality may cause numerically large deviations between the real-emission corrections
and the corresponding subtraction terms. Though the cancellation of infrared divergences still takes place,
the associated large weight fluctuations may significantly affect the convergence of the Monte-Carlo integration.
The problem becomes manifest when interfacing the fixed-order NLO calculation to a parton shower.
The difference in matrix-element weights arising from resonant propagators being shifted off resonance
by means of adding radiation and mapping momenta from Born to real-emission kinematics bears no relation
with the logarithms to be resummed by the parton shower, yet its numerical impact may be similar.
This motivates the usage of an improved kinematics mapping by means of pseudo-dipoles.

\subsection{Catani-Seymour pseudo-dipole formalism}\label{sec:originalIDformalism}
The concept of pseudo-dipoles was introduced in~\cite{Catani:1996vz} to cope with the situation where a subset of the
final-state partons lead to the production of identified hadrons. In such a scenario, both emitter and spectator of a
dipole may be ``identified'' in the sense that they fragment into identified hadrons. Because the directions of the identified
hadrons are measurable, neither emitter nor spectator parton in the dipole can be allowed to absorb the recoil when mapping the
momenta of the real-emission final state to a Born configuration.
Instead the kinematics is balanced by adjusting the momenta of all non-identified final-state particles (not just partons). 
This idea is reminiscent of standard dipoles with initial-state emitter and initial-state spectator. In fact,
pseudo-dipoles may be thought of as a generalization of these configurations.

In order to satisfy the constraint that both emitter and spectator retain their direction, an additional momentum
must be introduced that can absorb the recoil in the momentum mapping from real-emission to Born kinematics.
This auxiliary momentum is defined as
\begin{align}\label{eq:def_n}
n^\mu = p_\mathrm{in}^\mu - \sum_{\alpha \in \left\{\mathrm{id}\right\}} p_\alpha^\mu\;,
\end{align}
where $p_\mathrm{in}$ is the total incoming momentum in the process, and the sum runs over all outgoing identified particles.
The eventual dependence of the subtraction term on $n^\mu$ accounts for the term \textit{pseudo}-dipole.
An immediate consequence of this definition is that there are only two types of pseudo-dipoles, because the
spectator momentum is always in the final state. This is in contrast to standard Catani-Seymour dipoles,
which have four different types, corresponding to all combinations of initial-state or final-state emitter
with initial-state or final-state spectator. We denote pseudo-dipoles with final-state emitter as
$\mathcal{D}^{(n)}_{ai,b}$ and pseudo-dipoles with initial-state emitter as $\mathcal{D}^{(n)ai}_b$.

\subsubsection{Final-state singularities: differential form}\label{sec:org_fin_state_sing}
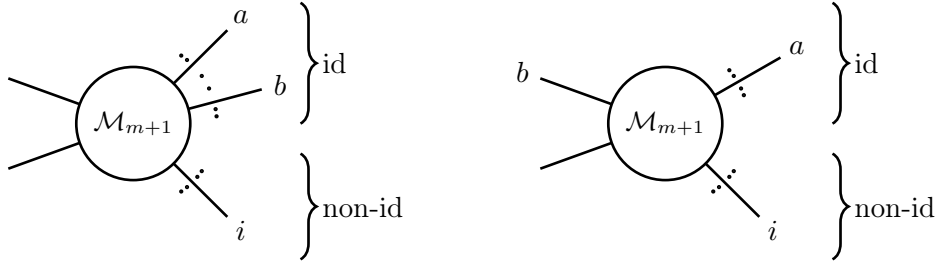
\begin{figure}[t]
\begin{center}
\begin{tikzpicture}[line width=1.0 pt, scale=1.0]
	\draw (0,0) circle (.75);
	\node at (0,0){\small{$\mathcal{M}_{m+1}$}};
	\draw (160:0.75)--(160:1.75);
	\draw (200:0.75)--(200:1.75);
	\draw[fill=black] (55:1.1) circle (.01);
	\draw[fill=black] (50:1.1) circle (.01);
	\draw (45:0.75)--(45:1.75);
	\node at (45:2){$a$};
	\draw[fill=black] (25:1.1) circle (.01);
	\draw[fill=black] (35:1.1) circle (.01);
	\draw (15:0.75)--(15:1.75);
	\node at (15:2){$b$};
	\draw[fill=black] (10:1.1) circle (.01);
	\draw[fill=black] (5:1.1) circle (.01);
	\draw (315:0.75)--(315:1.75);
	\draw[fill=black] (325:1.1) circle (.01);
	\draw[fill=black] (320:1.1) circle (.01);
	\node at (315:2){$i$};
	\draw[fill=black] (310:1.1) circle (.01);
	\draw[fill=black] (305:1.1) circle (.01);
	\draw [decorate,decoration={brace,amplitude=6pt}](2.2,1.6) -- (2.2,0.0) node [black,midway,label=right:id]{};
	\draw [decorate,decoration={brace,amplitude=6pt}](2.2,-0.4) -- (2.2,-1.8) node [black,midway,label=right:non-id]{};
\begin{scope}[shift={(7,0)}]
	\draw (0,0) circle (.75);
	\node at (0,0){\small{$\mathcal{M}_{m+1}$}};
	\draw (160:0.75)--(160:1.75);
	\node at (160:2){$b$};
	\draw (200:0.75)--(200:1.75);
	\draw[fill=black] (40:1.1) circle (.01);
	\draw[fill=black] (35:1.1) circle (.01);
	\draw (30:0.75)--(30:1.75);
	\node at (30:2){$a$};
	\draw[fill=black] (25:1.1) circle (.01);
	\draw[fill=black] (20:1.1) circle (.01);
	\draw (315:0.75)--(315:1.75);
	\draw[fill=black] (325:1.1) circle (.01);
	\draw[fill=black] (320:1.1) circle (.01);
	\node at (315:2){$i$};
	\draw[fill=black] (310:1.1) circle (.01);
	\draw[fill=black] (305:1.1) circle (.01);
	\draw [decorate,decoration={brace,amplitude=6pt}](2.2,1.6) -- (2.2,0.0) node [black,midway,label=right:id]{};
	\draw [decorate,decoration={brace,amplitude=6pt}](2.2,-0.4) -- (2.2,-1.8) node [black,midway,label=right:non-id]{};
\end{scope}
\end{tikzpicture}
\end{center}
\caption{Labeling for pseudo-dipoles with final-state singularity. 
  Left: Identified final-state spectator. 
  Right: Initial-state spectator.}\label{fig:fin_sing_assignment}
\end{figure}

The pseudo-dipole for splittings of final-state partons reads~\cite{Catani:1996vz}
\begin{align}\label{eq:pseudo_dipole_fs}
\mathcal{D}^{(n)}_{ai,b}  = -\frac{1}{2p_ap_i} 
\tensor[_{m,a,\hdots}]{\langle \hdots, \widetilde{ai}, \hdots, b, \hdots | \frac{\mathbf{T}_b \mathbf{T}_{ai}}{\mathbf{T}_{ai}^2} \mathbf{V}^{(n)}_{ai,b} | \hdots, \widetilde{a\imath}, \hdots, b, \hdots \rangle}{_{m,a,\hdots}}\;,
\end{align}
where $a$ refers to an identified final-state parton (the emitter) and the color spectator $b$ may either be
another identified final-state parton or an initial-state parton.
This situation is depicted in Fig.~\ref{fig:fin_sing_assignment}. 
The kinematics in the correlated Born matrix element is given as follows: The momentum of the emitter is scaled as
\begin{align}
\tilde{p}_{ai}^\mu = \frac{1}{z_{ain}}\, p_a^\mu 
\hspace{1cm}\mathrm{where}\hspace{1cm}
z_{ain} = \frac{p_a n}{(p_a+p_i)n}\label{eq:zain}\;.
\end{align}
All non-identified particles (not just partons) in the final state are Lorentz-transformed by
\begin{align}
\tilde{k}_j^\mu &= \Lambda^\mu_{\ \nu}(K,\tilde{K}) k_j^\nu \hspace{1.5cm}\mathrm{where}\\
\Lambda^\mu_{\ \nu}(K,\tilde{K}) &= g^\mu_{\ \nu} - \frac{2(K+\tilde{K})^\mu (K+\tilde{K})_\nu}{(K+\tilde{K})^2} + \frac{2\tilde{K}^\mu K_\nu}{K^2}\;.\label{eq:Lorentztrafo}
\end{align}
The momenta $K^\mu$ and $\tilde{K}^\mu$ are given by
\begin{align}
K^\mu &= n^\mu - p_i^\mu \\
\tilde{K}^\mu &= n^\mu - (1-x_{ain})p_a^\mu 
\hspace{1cm}\mathrm{where}\hspace{1cm}
x_{ain} = \frac{(p_a-p_i)n}{p_a n}\;.\label{eq:xain}
\end{align}
The remaining momenta, namely those of identified and initial-state particles (in particular of the color spectator: parton $b$) remain unchanged.

The complete list of pseudo-dipole insertion operators is given in~\cite{Catani:1996vz}.
As we focus on processes with no final-state gluons at Born level, the only one
relevant to our computations is the $q \to qg$ insertion operator, which reads
\begin{align}
\left\langle s \left| \mathbf{V}_{q_a g_i,b}^{(n)} \right| s^\prime\right\rangle &= 8\pi \mu^{2\epsilon} \alpha_s C_F \left[ 2\, \frac{v_{i,ab}}{z_{ain}} - (1+z_{ain}) - \epsilon(1-z_{ain}) \right] \delta_{ss^\prime}\;,\label{eq:VIDqg}
\end{align}
where 
\begin{align}
v_{i,ab}=\frac{p_ap_b}{p_i(p_a+p_b)}\;.\label{eq:v_iab}
\end{align}
The integrated splitting kernel is defined as
\begin{align}
  \frac{\alpha_s}{2\pi}\frac{1}{\Gamma(1-\epsilon)} \left( \frac{4\pi\mu^2}{2p_ap_b} \right)^\epsilon \bar{\mathcal{V}}_{ai,a}% (z_{ain};\epsilon;p_a,p_b,n)
  := \int \left[ \mathrm{d}p_i(n,p_a,z) \right] \frac{1}{2p_ap_i} \langle \mathbf{V}_{ai,b}^{(n)}(z_{ain};v_{iab}) \rangle.
\end{align}
where $\left[ \mathrm{d}p_i(n,p_a,z) \right]$ is the one-emission phase-space differential
obtained by factorizing the real-emission phase space. One obtains
\begin{align}
\bar{\mathcal{V}}_{q,q}(z;\epsilon;p_a,p_b,n) =&\; -\frac{1}{\epsilon}P^{qq}(z) + \delta(1-z)V_{qg}(\epsilon) + \tilde{K}^{qq}(z) + \bar{K}^{qq}(z) + P^{qq}(z) \mathrm{ln}z\nonumber\\
&\; + \mathcal{L}^{q,q}(z;p_a,p_b,n) + \mathcal{O}(\epsilon),\label{eq:Vbar}
\end{align}
where $V_{qg}(\epsilon)$ comprises all singularities needed to cancel the poles present in the virtual corrections.
It is given in Eq.~(5.32) of~\cite{Catani:1996vz}, and all other functions are listed in Appendix C of~\cite{Catani:1996vz}.
The single poles proportional to the Altarelli-Parisi splitting functions do not appear in the integrated standard CS-dipole terms. They are canceled by collinear mass factorization counterterms, which -- in our approach --
can be viewed as the one-loop contribution to the partonic fragmentation function to be convoluted with the
subtracted hard cross section.

\subsubsection{Initial-state singularities: differential form}\label{sec:org_in_state_sing}
\begin{figure}[t]
\begin{center}
\begin{tikzpicture}[line width=1.0 pt, scale=1.0]
	\draw (0,0) circle (.75);
	\node at (0,0){\small{$\mathcal{M}_{m+1}$}};
	\draw (160:0.75)--(160:1.75);
	\node at (160:2){$a$};
	\draw (200:0.75)--(200:1.75);
	\draw[fill=black] (40:1.1) circle (.01);
	\draw[fill=black] (35:1.1) circle (.01);
	\draw (30:0.75)--(30:1.75);
	\node at (30:2){$b$};
	\draw[fill=black] (25:1.1) circle (.01);
	\draw[fill=black] (20:1.1) circle (.01);
	\draw (315:0.75)--(315:1.75);
	\draw[fill=black] (325:1.1) circle (.01);
	\draw[fill=black] (320:1.1) circle (.01);
	\node at (315:2){$i$};
	\draw[fill=black] (310:1.1) circle (.01);
	\draw[fill=black] (305:1.1) circle (.01);
	\draw [decorate,decoration={brace,amplitude=6pt}](2.2,1.6) -- (2.2,0.0) node [black,midway,label=right:id]{};
	\draw [decorate,decoration={brace,amplitude=6pt}](2.2,-0.4) -- (2.2,-1.8) node [black,midway,label=right:non-id]{};
\begin{scope}[shift={(7,0)}]
	\draw (0,0) circle (.75);
	\node at (0,0){\small{$\mathcal{M}_{m+1}$}};
	\draw (160:0.75)--(160:1.75);
	\node at (160:2){$a$};
	\draw (200:0.75)--(200:1.75);
	\node at (200:2){$b$};
	\draw[fill=black] (40:1.1) circle (.01);
	\draw[fill=black] (35:1.1) circle (.01);
	\draw (30:0.75)--(30:1.75);
	\draw[fill=black] (25:1.1) circle (.01);
	\draw[fill=black] (20:1.1) circle (.01);
	\draw (315:0.75)--(315:1.75);
	\draw[fill=black] (325:1.1) circle (.01);
	\draw[fill=black] (320:1.1) circle (.01);
	\node at (315:2){$i$};
	\draw[fill=black] (310:1.1) circle (.01);
	\draw[fill=black] (305:1.1) circle (.01);
	\draw [decorate,decoration={brace,amplitude=6pt}](2.2,1.6) -- (2.2,0.0) node [black,midway,label=right:id]{};
	\draw [decorate,decoration={brace,amplitude=6pt}](2.2,-0.4) -- (2.2,-1.8) node [black,midway,label=right:non-id]{};
\end{scope}
\end{tikzpicture}
\end{center}
\caption{Labeling for pseudo-dipoles with initial-state singularity. Left: Identified final-state spectator. Right: Initial-state spectator }\label{fig:in_sing_assignment}
\end{figure}
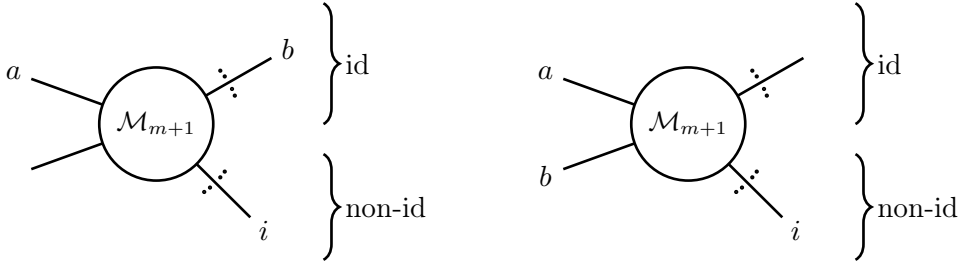

The pseudo-dipole with an initial-state emitter and identified final-state or initial-state spectator reads
\begin{align}\label{eq:pseudo_dipole_is}
\mathcal{D}_{b}^{(n)ai} = -\frac{1}{2p_ap_i}\frac{1}{x_{ain}}
\tensor[_{m,a,\hdots}]{\left\langle \hdots,  \widetilde{a\imath}, \hdots, b, \hdots \right| \frac{\mathbf{T}_b \mathbf{T}_{ai}}{\mathbf{T}_{ai}^2} \mathbf{V}_{b}^{(n)ai} \left| \hdots,  \widetilde{a\imath}, \hdots, b, \hdots \right\rangle}{_{m}}.
\end{align}
The two cases are sketched in Fig.~\ref{fig:in_sing_assignment}.
The momentum mapping is chosen as follows:
\begin{align}
  %&\mathrm{emitter:} &
  & \tilde{p}_{ai}^\mu = x_{ain} p_a^\mu,
  %&\mathrm{spectators:}
  && \tilde{k}_j^\mu = \Lambda^\mu_{\ \nu} k_j^\mu .
\end{align}
The momenta of all identified final-state partons %$p_\alpha^\mu$ 
and the other initial-state parton %$p_b^\mu$ 
are left unchanged. 
The Lorentz transformation matrix is the same as in Eq.~\eqref{eq:Lorentztrafo}
and $x_{ain}$ is defined in Eq.~\eqref{eq:xain}.
The splitting functions are given by
\begin{align}\label{eq:VIDinitial}
\left\langle s \left| \mathbf{V}_{b}^{(n)q_a g_i} \right| s^\prime\right\rangle &= 8\pi \mu^{2\epsilon} \alpha_s C_F \left[ 2 v_{i,ab} - (1+x_{ain}) - \epsilon(1-x_{ain}) \right] \delta_{ss^\prime},\\%\label{eq:VIDinitial_q_qg}\\
\left\langle s \left| \mathbf{V}_{b}^{(n)g_a \bar{q}_i} \right| s^\prime\right\rangle &= 8\pi \mu^{2\epsilon} \alpha_s T_R \left[ 1-\epsilon-2x_{ain}(1-x_{ain}) \right]\delta_{ss^\prime},\nonumber\\%\label{eq:VIDinitial_g_qq}\\
\left\langle \mu \left| \mathbf{V}_{b}^{(n)q_a q_i} \right| \nu\right\rangle &= 8\pi \mu^{2\epsilon} \alpha_s C_F \left[ -g^{\mu\nu}x_{ain} + \frac{1-x_{ain}}{x_{ain}} \frac{4p_ip_a}{2(p_an)(p_in)-n^2p_ip_a} \right.\nonumber\\
&\ \left. \left( \frac{np_a}{p_ip_a}p_i^\mu - n^\mu \right)\left( \frac{np_a}{p_ip_a}p_i^\nu - n^\nu \right) \right],\nonumber\\%\label{eq:VIDinitial_q_gq}\\
\left\langle \mu \left| \mathbf{V}_{b}^{(n)g_a g_i} \right| \nu\right\rangle &= 16\pi \mu^{2\epsilon} \alpha_s C_A \left[ -g^{\mu\nu}\left(v_{i,ab}-1 + x_{ain}(1-x_{ain})\right) +(1-\epsilon) \frac{1-x_{ain}}{x_{ain}}\right.\nonumber\\
&\ \left.  \frac{4p_ip_a}{2(p_an)(p_in)-n^2p_ip_a}\left( \frac{np_a}{p_ip_a}p_i^\mu - n^\mu \right)\left( \frac{np_a}{p_ip_a}p_i^\nu - n^\nu \right) \right],\nonumber%\label{eq:VIDinitial_g_gg}
\end{align}
The integrated splitting functions are defined by:
\begin{align}
\frac{\alpha_s}{2\pi}\frac{1}{\Gamma(1-\epsilon)} \left( \frac{4\pi\mu^2}{2p_ap_b} \right)^\epsilon \tilde{\mathcal{V}}^{a,ai} (x;\epsilon;p_a,p_b,n) := \int \left[ \mathrm{d}p_i(n,p_a,x) \right] \frac{1}{2p_ap_i}\frac{n_s(\widetilde{a\imath})}{n_s(a)} \langle \mathbf{V}_b^{(n)ai}(x;v_{iab}) \rangle,
\end{align}
where $n_s(a)$ is the number of polarisations of parton $a$.
The result is identical to the final-state case, i.e. Eq.~\eqref{eq:Vbar}:
\begin{align}
\tilde{\mathcal{V}}^{a,b} (x;\epsilon;p_a,p_b,n) = \bar{\mathcal{V}}^{a,b} (x;\epsilon;p_a,p_b,n).
\end{align}

\subsection{Application to resonance-aware subtraction}\label{sec:exploitPseudoDipoles} 
\begin{figure}[t]
\begin{center}
\begin{tikzpicture}[line width=1.0 pt, scale=1.3,
        line/.style={dotted,thick,draw=black, postaction={decorate},
        decoration={markings,
        mark=at position 0.99 with {\arrow[draw=black]{>}}}},
        noline/.style={dotted,thick,draw=white, postaction={decorate,
        decoration={markings,
        mark=at position 0.99 with {\arrow[draw=black]{>}}}}},
        linenoarrow/.style={dotted,thick,draw=black}]
	\draw[gluon] (-1,1)--(0,0);
	\draw[gluon] (-1,-1)--(0,0);
	\draw[gluon] (-0.5,-0.5)--(0.3,-1);
	\node at (0.6,-1) {$g$};
	\draw[gluon] (0,0)--(1,0);
	\draw[fermionbar] (1,0)--node[label=above:$\bar{t}$]{}(1.5,0.5);
	\draw[fermionbar] (1.5,0.5)--(2,1);
	\draw[vector] (1.5,0.5)--(2,0.3);
	\draw[fermion] (1,0)--node[label=below:$t$]{}(1.5,-0.5);
	\draw[fermion] (1.5,-0.5)--(2,-1);
	\draw[vector] (1.5,-0.5)--(2,-0.3);
	\node at (2.3,1) {$\bar{b}$};
	\node at (2.4,0.3) {$W^-$};
	\node at (2.4,-0.3) {$W^+$};
	\node at (2.3,-1) {$b$};

\begin{scope}[shift={(3.5,0)}]
	\node at (1,0.2){$\mathrm{ID-momenta}$};
	\draw[feynarrow] (0,0)--(2.0,0);
	\node at (1,-0.2){$\mathrm{mapping}$};
\end{scope}

\begin{scope}[shift={(7,0)}]
	\draw[gluon] (-1,0.5)--(-0.5,0);
	\draw[gluon] (-1,-0.5)--(-0.5,0);	
	\draw[gluon] (-0.5,0)--(0.5,0); 
	\draw[fermionbar] (0.5,0)--node[label=above:$\bar{t}$]{}(1,0.5);
	\draw[fermionbar] (1,0.5)--(1.5,1);
	\draw[fermionbar] (1.5,1)--(2,1.5);
	\draw[vector] (1,0.5)--(2,0.3);
	\draw[fermion] (0.5,0)--node[label=below:$t$]{}(1,-0.5);
	\draw[fermion] (1,-0.5)--(2,-1.5);
	\draw[vector] (1,-0.5)--(2,-0.3);
	\node at (2.3,1.5) {$\bar{b}$};
	\node at (2.4,0.3) {$W^-$};
	\node at (2.4,-0.3) {$W^+$};
	\node at (2.3,-1.5) {$b$};
	\draw[line] plot [smooth] coordinates {(2.0,-1.3) (1.4,-0.7)
	  (2.0,-0.5)};
	\draw[noline] plot [smooth] coordinates { (2.0,-0.5)(1.4,-0.7)
	  (2.0,-1.3)};
	\draw[line] plot [smooth] coordinates {(2.0,1.3) (1.4,0.7)
	  (2.0,0.5)};
	\draw[noline] plot [smooth] coordinates { (2.0,0.5)(1.4,0.7)
	  (2.0,1.3)};
	\draw[linenoarrow] plot [smooth] coordinates {(1.4,0.7)
	 (0.8,0.0) (1.4,-0.7)};
	\draw[linenoarrow] plot [smooth] coordinates {(-1,-0.8)(-0.4,-0.2)(0.4,-0.2)
	 (0.8,0.0)};
\end{scope}
\end{tikzpicture}
\end{center}
\caption{Exemplary momentum mapping for $gg \to bW^+\bar{b}W^-$ from a real configuration to an underlying Born configuration for an initial-state singularity.\newline
The curved arrow on the right indicates the flow of the recoil.}\label{fig:assignment_IDexample_pp}
\end{figure}
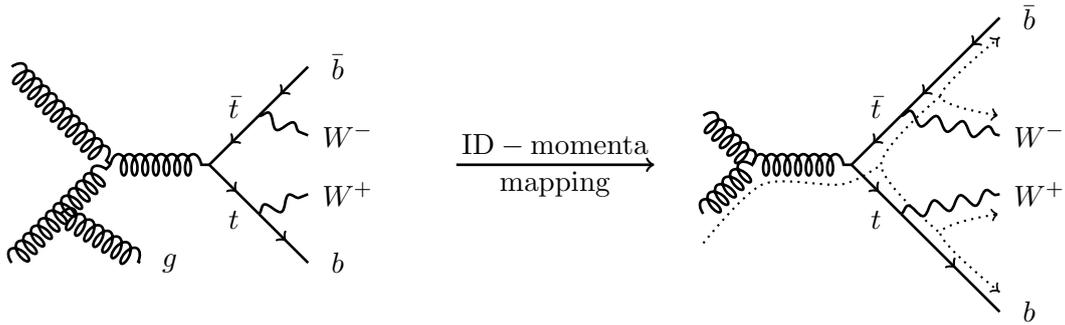

We will now explain how to apply pseudo-dipoles to resonant processes,
starting with the simplest case of processes without hadronic initial states.
Pseudo-dipole subtraction terms for final-state singularities that preserve the invariant mass of intermediate resonances
can be constructed using the following algorithm:
\begin{enumerate}
\item If the emitter is the decay product of a resonance and the spectator is not
  a decay product of the same resonance, the dipole is replaced by a pseudo-dipole
  where the emitter and all particles except for the emission and the remaining
  decay products of the resonance are identified.
\item If the emitter is not a decay product of a resonance but the spectator
is, the dipole is replaced by a pseudo-dipole
  where the emitter and all particles, which are not decay products of the resonance to which the spectator
  belongs are identified. 
\item If emitter and spectator are decay products of the same resonance,
  the standard CS-subtraction formalism is used.
\end{enumerate}
It is clear that these rules can only be applied unambiguously once the diagrammatic structure
of the real-emission corrections is simple enough for a clear assignment of ``decay products''
to be made. Despite this restriction, the algorithm can be used in a variety of processes,
among them the highly relevant example of top-quark pair production, both at lepton and at hadron colliders.
%(when including the rules below).
\begin{figure}[t]
\begin{center}
\begin{tikzpicture}[line width=1.0 pt, scale=1.3,
        line/.style={dotted,thick,draw=black, postaction={decorate},
        decoration={markings,
        mark=at position 0.99 with {\arrow[draw=black]{>}}}},
        noline/.style={dotted,thick,draw=white, postaction={decorate,
        decoration={markings,
        mark=at position 0.99 with {\arrow[draw=black]{>}}}}},
        linenoarrow/.style={dotted,thick,draw=black}]
	\draw[fermion] (-1,0.5)--(-0.5,0);
	\draw[fermionbar] (-1,-0.5)--(-0.5,0);	
	\draw[vector] (-0.5,0)--(0.5,0); 
	\draw[fermionbar] (0.5,0)--node[label=above:$\bar{t}$]{}(1,0.5);
	\draw[fermionbar] (1,0.5)--(1.5,1);
	\draw[fermionbar] (1.5,1)--(2,1.5);
	\draw[gluon] (1.5,1)--(2,0.8);
	\draw[vector] (1,0.5)--(2,0.3);
	\draw[fermion] (0.5,0)--node[label=below:$t$]{}(1,-0.5);
	\draw[fermion] (1,-0.5)--(2,-1.5);
	\draw[vector] (1,-0.5)--(2,-0.3);
	\node at (2.3,1.5) {$\bar{b}$};
	\node at (2.3,0.8) {$g$};
	\node at (2.4,0.3) {$W^-$};
	\node at (2.4,-0.3) {$W^+$};
	\node at (2.3,-1.5) {$b$};
\begin{scope}[shift={(3.5,0)}]
	\node at (1,0.2){$\mathrm{ID-momenta}$};
	\draw[feynarrow] (0,0)--(2.0,0);
	\node at (1,-0.2){$\mathrm{mapping}$};
\end{scope}
\begin{scope}[shift={(7,0)}]
	\draw[fermion] (-1,0.5)--(-0.5,0);
	\draw[fermionbar] (-1,-0.5)--(-0.5,0);	
	\draw[vector] (-0.5,0)--(0.5,0); 
	\draw[fermionbar] (0.5,0)--node[label=above:$\bar{t}$]{}(1,0.5);
	\draw[fermionbar] (1,0.5)--(1.5,1);
	\draw[fermionbar] (1.5,1)--(2,1.5);
	\draw[vector] (1,0.5)--(2,0.3);
	\draw[fermion] (0.5,0)--node[label=below:$t$]{}(1,-0.5);
	\draw[fermion] (1,-0.5)--(2,-1.5);
	\draw[vector] (1,-0.5)--(2,-0.3);
	\node at (2.3,1.5) {$\bar{b}$};
	\node at (2.4,0.3) {$W^-$};
	\node at (2.4,-0.3) {$W^+$};
	\node at (2.3,-1.5) {$b$};
	\draw[line] plot [smooth] coordinates { (2.0,1.3) (1.4,0.7)
	  (2.0,0.5)};
\end{scope}
\end{tikzpicture}
\end{center}
\caption{Possible real correction configuration for $W^+W^-b\bar{b}$ production and Born configuration of associated pseudo-dipoles. 
The curved arrow on the right indicates the flow of the recoil.  
}\label{fig:assignment_IDexample_ee}
\end{figure}
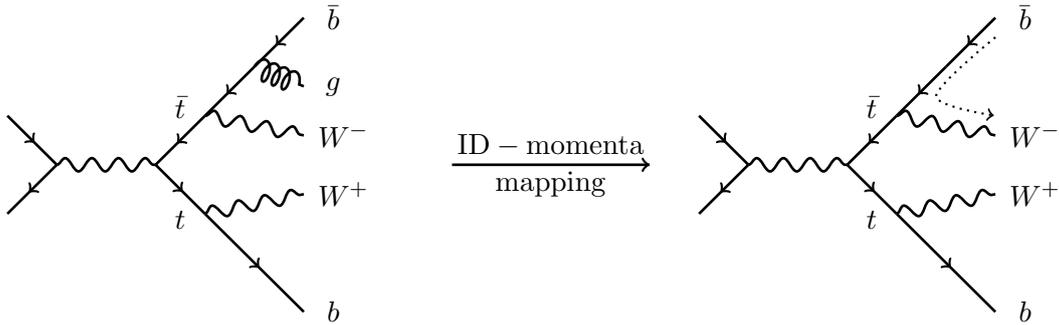
Consider again the example $e^+e^- \to W^+W^-b\bar{b}$. If $\sqrt{s} > 2m_t$ the dominant
contribution to the cross section stems from diagrams like the one on the left-hand side of
Fig.~\ref{fig:assignment_CSexample}. The standard CS-dipole to cover the soft and collinear singularity
associated to this diagram is constructed by using the Born-level diagram on the right-hand side
of Fig.~\ref{fig:assignment_CSexample}. In this situation the recoil from the emitter parton
$\bar{b}$ to the spectator parton $b$ affects the potentially resonant top quark propagators.
To avoid this, we replace by means of the above algorithm the standard CS-dipole by a pseudo-dipole
and formally ``identify'' particles. As the first rule takes precedence we identify $\bar{b}$,
$W^+$ and $b$. In this manner, the $W^-$ boson is the only particle left to absorb the recoil. 
Hence the momentum of the top-quarks are unaltered and we have achieved our aim.
The momentum flow corresponding to this situation is depicted in Fig.~\ref{fig:assignment_IDexample_ee}.
The same reasoning is applied to the pseudo-dipole in which the $b$ quark is the emitter.
\\
When considering initial-state partons, there are in principle three more dipoles types, namely those with
final-state emitter and initial-state spectator (FI), initial-state emitter and final-state spectator (IF) and finally initial-state emitter and spectator (II).
As already alluded to in sec~\ref{sec:originalIDformalism}, FI-pseudo-dipoles are the same as FF-pseudo-dipoles.
Hence, the rules from above apply.\footnote{In order to be applicable to initial-state spectators, we have to rephrase the second rule to:
If the emitter is not the decay product of a resonance and the spectator is an initial-state parton, the dipole is replaced by a pseudo-dipole where the emitter and all particles, which are not decay products of any resonance are identified.}
The case of IF-pseudo-dipoles can be generally treated by the following rule:
\begin{enumerate}
\item If the emitter is in the inital state and the spectator is the decay product of a resonance, the dipole is replaced by a pseudo-dipole where no particle is identified.
\end{enumerate}
Thus the recoil is the same as for standard II-CS-dipoles.
For II-pseudo-dipoles, we choose the same recoil scheme as for IF-pseudo-dipoles.
In doing so, pseudo- and standard CS-dipoles of II-type are identical, which stresses the fact that standard II-CS-dipoles are already resonance-aware.
This is because the momentum mapping is given in the form of a Lorentz transformation, which preserves the virtuality of all final-state particles and thus all intermediate resonances, hence rendering the mapping resonance-aware.
Figure~\ref{fig:assignment_IDexample_pp} displays the recoil flow for pseudo-dipoles with initial-state singularities at the example of $gg \to W^+W^-gb\bar{b}$.

In view of our application example $pp \to W^+W^-j_bj_b$ in Sec.~\ref{sec:ee_bwbw}, we like to comment on the situation in which one has to deal with multiple (sub-)processes.
In this concrete example we replace standard CS- with pseudo-dipoles only if both real and underlying Born-configuration comprise at least one $b$- and one $\bar{b}$-quark in the final state.
This together with the rules outlined above renders the use of pseudo-dipoles for this process unambiguous.
% \todo{Seb: There is no ambiguity in $Gb \to W^+W^-bb\bar{b}$. Explain exactly how this works! (will be done by Seb)}
% Obvious for initial-state singularities.
% For final-state singularity, pseudo-dipoles are not used, since the underlying Born-ME does never have one b- and one bbar-quark.

We stress at this point two vital differences of the use of pseudo-dipoles for resonance-aware subtraction and their original purpose:
\begin{itemize}
\item We do not identify particles throughout the calculation, but identify different particles depending on the subtraction term.
\item We integrate over the momenta of the identified partons by means of adding partonic fragmentation functions.
\end{itemize}

We will show in the following how this affects the {\bf H}-, {\bf K}- and {\bf P}-terms given in~\cite{Catani:1996vz}, by focusing on the case of final-state singularities in some detail before we quote the analogous results for initial-state singularities.

\subsubsection{Final-state singularities: integrated form}\label{sec:fin_sing_integrated}
For simplicity, we first consider a configuration with no initial-state partons and $m$ final-state
(anti-)quarks at Born level. In the following, the integration over non-QCD particles
shall be understood whenever we write $\int_m \mathrm{d}\phi_m$.

For now we will not integrate over the momentum of the merged parton $\widetilde{a\imath}$.
We denote this with a subscript $-1$ at the integral sign. The integral over the pseudo-dipoles is then given by
\begin{align}
\int_{m+1-1}\mathrm{d}\sigma^\mathrm{S}\Big|_{\mathrm{FF}}^\mathrm{ID} =& \ \mathcal{N}_{in}\sum_{\left\{m+1\right\}} \int_{m+1-1} \mathrm{d}\phi_{m+1-1} \frac{1}{S_{\left\{m+1\right\}}} \sum_{\substack{\mathrm{pairs} \\ a,i}} \sum_{b \neq a,i} \mathcal{D}_{ai,b}^{(n)}F^{(m)}_J \bigg|_{\substack{a \in \left\{\mathrm{fin}\right\} \\ b \in \left\{\mathrm{fin}\right\}}} \nonumber\\
=& -\mathcal{N}_{in} \int_0^1 \frac{\mathrm{d}z}{z^{2-2\epsilon}} \sum_{\widetilde{a\imath}=1}^m \sum_{\left\{m\right\}} \int_{m-1} \mathrm{d}\phi_{m-1} \frac{1}{S_{\left\{m\right\}}} F^{(m)}_J \sum_{\substack{b=1 \\ b \neq \widetilde{a\imath} }}^m \left| \mathcal{M}^{\widetilde{a\imath}~b}_m\right|^2 \nonumber\\
&\quad\times \frac{\alpha_s}{2\pi} \frac{1}{\Gamma(1-\epsilon)} \left( \frac{4\pi\mu^2}{2p_ap_b} \right)^\epsilon \frac{1}{\mathbf{T}_{\widetilde{a\imath}}^2}   \bar{\mathcal{V}}_{\widetilde{a\imath},a}(z;\epsilon;p_a,p_b,n) \delta_{\widetilde{a\imath}~a}\bigg|_{\substack{\widetilde{a\imath} \in \left\{\mathrm{fin}\right\} \\ b \in \left\{\mathrm{fin}\right\}}}. \label{eq:IDsub}
\end{align}
Here $z$ is used as a shorthand notation for $z_{ain}$, given in Eq.~\eqref{eq:zain}.
As we only consider $\widetilde{a\imath} \to a+i$ as $q \to q+g$ splittings, $\widetilde{a\imath}$ and $a$ are always quarks. 
The corresponding integrated splitting function is given in Eq.~\eqref{eq:Vbar}.
Each identified final-state parton contributes a collinear mass factorization counterterm
%\begin{equation}
%  \int_{m-1}\mathrm{d}\sigma^\mathrm{C} =& \sum_{a=1}^m \int_{m-1}\mathrm{d}\sigma^\mathrm{C}_{a}(p_a)\label{eq:sumIDcoll}
%\end{equation}
%where
\begin{align}
  \int_{m-1}\mathrm{d}\sigma^\mathrm{C}_{a}(p_a)\Big|_{\mathrm{FF}}^\mathrm{ID} =&\; - \frac{\alpha_s}{2\pi}\frac{1}{\Gamma(1-\epsilon)} \int_0^1 \frac{\mathrm{d}z}{z^{2-2\epsilon}}\;  \delta_{\widetilde{a\imath}~a}\label{eq:IDcoll}\\
  &\quad\times \left.\left[ -\frac{1}{\epsilon}\left(\frac{4\pi\mu^2}{\mu_F^2}\right)^\epsilon P_{\widetilde{a\imath}~a}(z) + H^{F.S.}_{\widetilde{a\imath}~a}(z)\right] \int_{m-1}\mathrm{d}\sigma^\mathrm{B}_{\widetilde{a\imath}}\left(\frac{p_a}{z}\right)\right|_{\widetilde{a\imath} \in \left\{\mathrm{fin}\right\}} \;.\nonumber
\end{align}
where the Born cross-section is given by
\begin{align}
\int_m\mathrm{d}\sigma^\mathrm{B} = \mathcal{N}_{in} \sum_{\left\{m\right\}} \int_m \mathrm{d}\phi_{m} \frac{1}{S_{\left\{m\right\}}} F^{(m)}_J \left| \mathcal{M}_m\right|^2.\label{eq:Born}
\end{align}
To avoid cluttering the notation, we drop the sub- and superscripts indicating
pseudo-dipoles with final-state emitter and (color-)spectator.
When we add Eq.~\eqref{eq:IDsub} and sum Eq.~\eqref{eq:IDcoll} for all identified partons,
we can define an insertion operator:
\begin{align}
\sum_{\widetilde{a\imath}=1}^m \sigma^\mathrm{I}(p_a) := \int_{m+1-1}\mathrm{d}\sigma^\mathrm{S} + \sum_{a=1}^m\int_{m-1}\mathrm{d}\sigma^\mathrm{C}_a =: \sum_{\widetilde{a\imath}=1}^m \int_0^1 \frac{\mathrm{d}z}{z^{2-2\epsilon}} \int_{m-1} \mathrm{d}\sigma^B \left( \frac{p_a}{z} \right) \cdot \mathbf{\hat{I}}_{\widetilde{a\imath}}\label{eq:my_insert}
\end{align}
where $\mathrm{d}\sigma^B\left(\frac{p_a}{z}\right) \cdot \mathbf{\hat{I}}_{\widetilde{a\imath}}$ indicates
that the squared Born matrix element %$\left| \mathcal{M}_m \right|^2 = \tensor[_{m}]{\left\langle p_1, \hdots, p_m | p_1, \hdots, p_m \right\rangle}{_{m}}$
in Eq.~\eqref{eq:Born} is replaced by the spin- and color-correlated Born matrix element.
%\begin{align}
%\tensor[_{m}]{\left\langle p_1, \hdots, \frac{p_a}{z}, \hdots, p_m \right| \hat{\mathbf{I}}_{\widetilde{a\imath}} \left| p_1, \hdots, \frac{p_a}{z}, \hdots, p_m \right\rangle}{_{m}}.
%\end{align}
Note that the parton $\widetilde{a\imath}$ in this expression carries momentum $\tilde{p}_{ai} = p_a/z$.
The insertion operator is given by
\begin{align}
\hat{\mathbf{I}}_{\widetilde{a\imath}} =& -\frac{\alpha_s}{2\pi}\frac{1}{\Gamma(1-\epsilon)}\Bigg\{ \sum_{\substack{b=1 \\ b \neq \widetilde{a\imath}}}^m \frac{\mathbf{T}_{\widetilde{a\imath}}\mathbf{T}_b}{\mathbf{T}_{\widetilde{a\imath}}^2} \left( \frac{4\pi\mu^2}{2p_ap_b} \right)^\epsilon  \delta_{\widetilde{a\imath}~a}\bar{\mathcal{V}}_{q,q}(z;\epsilon;p_a,p_b,n) \nonumber\\
&\quad+ \delta_{\widetilde{a\imath}~a} \left[ -\frac{1}{\epsilon}\left(\frac{4\pi\mu^2}{\mu_F^2}\right)^\epsilon P_{\widetilde{a\imath}~a}(z) + H^{F.S.}_{\widetilde{a\imath}~a}(z)\right]\Bigg\}.
\end{align}
From this expression we extract the insertion operator $\mathbf{I}_{\widetilde{a\imath}}$,
which comprises all singularities and is identical to the one for standard FF-dipoles:
\begin{align}
\mathbf{I}_{\widetilde{a\imath}}\left( p_1,\hdots,p_a,\hdots,p_m;\epsilon \right) =& -\frac{\alpha_s}{2\pi}\frac{1}{\Gamma(1-\epsilon)} \sum_{\substack{b=1 \\ b \neq \widetilde{a\imath}}}^m \frac{\mathbf{T}_{\widetilde{a\imath}}\mathbf{T}_b}{\mathbf{T}_{\widetilde{a\imath}}^2} \left( \frac{4\pi\mu^2}{2p_ap_b} \right)^\epsilon \mathcal{V}_{qg}(\epsilon). \label{eq:myI}
\end{align}
% Problem with prefactor comprising $p_a$ and not $\tilde{p}_{ai}$ is not there, because the whole insertion operator is multiplied with $\delta(1-z)$.
We split the remainder into $\mathbf{H}$- and $\mathbf{P}$-operators as
\begin{align}
\hat{\mathbf{I}}_{\widetilde{a\imath}} = \delta(1-z)\mathbf{I}_{\widetilde{a\imath}} + \mathbf{H}_{\widetilde{a\imath}} + \mathbf{P}_{\widetilde{a\imath}}.
\end{align}
In order to combine the $\mu_F$-dependent terms with the collinear counterterms, we used the identity
$\sum_{\substack{b=1 \\ b \neq \widetilde{a\imath}}}^m \mathbf{T}_b = -\mathbf{T}_{\widetilde{a\imath}}$,
which arises from color conservation~\cite{Catani:1996vz}. 
%The factor two in front of $P_{\widetilde{a\imath}~a}(z)\mathrm{ln}z$ in $\mathbf{H}$ is due to the shift of $z$ in the argument of the logarithm in $\mathbf{P}_{\widetilde{a\imath}}$.
After a few steps, we find up to $\mathcal{O}(\epsilon)$:
\begin{align}
\mathbf{P}_{\widetilde{a\imath}}\left( p_1,\hdots,\frac{p_a}{z},\hdots,p_m;z;\mu_F \right) =& \frac{\alpha_s}{2\pi}  \sum_{\substack{b=1 \\ b \neq \widetilde{a\imath}}}^m \frac{\mathbf{T}_{\widetilde{a\imath}}\mathbf{T}_b}{\mathbf{T}_{\widetilde{a\imath}}^2} \mathrm{ln} \frac{z \mu_F^2}{2p_ap_b} \delta_{\widetilde{a\imath}~a}P_{\widetilde{a\imath}~a}(z) 
\end{align}
and
\begin{align}
  &\hspace*{-1cm}\mathbf{H}_{\widetilde{a\imath}}\left( p_1,\hdots,p_a,\hdots,p_m;n;z \right) =\nonumber\\
  & -\frac{\alpha_s}{2\pi}\Bigg\{ \sum_{\substack{b=1 \\ b \neq \widetilde{a\imath}}}^m \frac{\mathbf{T}_{\widetilde{a\imath}}\mathbf{T}_b}{\mathbf{T}_{\widetilde{a\imath}}^2} \delta_{\widetilde{a\imath}~a} \Big[ \bar{K}^{\widetilde{a\imath}~a}(z) + 2P_{\widetilde{a\imath}~a}(z)\mathrm{ln}z + \tilde{K}^{\widetilde{a\imath}~a}(z) \nonumber\\
&\qquad\qquad+\mathcal{L}^{\widetilde{a\imath}~a}(z;p_a,p_b,n)\Big] + \delta_{\widetilde{a\imath}~a} H^{F.S.}_{\widetilde{a\imath}~a}(z) \Bigg\}.\label{eq:myH}
\end{align}
%\subsubsection*{Integrating over the Momentum of the Emitting Parton $\widetilde{a\imath}$}
In contrast to the original pseudo-dipole approach, we now replace the integration over
$p_a$ by an integration over the Born momentum $\tilde{p}_{ai}=p_a/z$ (see e.g.\ Ref.~\cite{Gleisberg:2007md}, p.24).
This leads to the following transformation:% and altered integration boundaries:
\begin{align}
\int \frac{\mathrm{d}^{D-1} p_a}{2|\vec{p}_a|} = 
%\int_0^P |\vec{q}_a|^2 \mathrm{d}|\vec{p}_a| \mathrm{d}\Omega = \int_0^{z\tilde{P}} z^3|\tilde{\vec{p}}_a|^2\mathrm{d}|\tilde{\vec{p}}_a|\mathrm{d}\Omega = 
z^{2-2\epsilon} \int \frac{\mathrm{d}^{D-1} \tilde{p}_a}{2|\vec{\tilde{p}}_a|}\;.
\end{align}
The Jacobian of the transformation cancels the prefactor $1/z^{2-2\epsilon}$ in Eq.~\eqref{eq:my_insert},
and we obtain
%where the upper integration bound $z\tilde{Q}$ is reduced by $z$ in comparison to every other parton exiting the matrix element.
\begin{align}
\sum_{\widetilde{a\imath}=1}^m \int \frac{\mathrm{d}^D p_a}{(2\pi)^{D-1}}\delta(p_a^2)\, \sigma^{\mathrm{I}}(p_a)
=& \sum_{\widetilde{a\imath}=1}^m \int_m \mathrm{d}\sigma^\mathrm{B}(p_1, \hdots, p_m)
 \int_0^1 \mathrm{d}z \Big[ \delta(1-z)\mathbf{I}_{\widetilde{a\imath}} + \mathbf{H}_{\widetilde{a\imath}}+\mathbf{P}_{\widetilde{a\imath}} \Big]\;.
\end{align}
The momentum of parton $\widetilde{a\imath}$ is $\tilde{p}_{ai}$ and thus no longer $z$-dependent.
This allows to simplify the operators. Setting $H^{F.S.}_{qq}(z) = 0$ (which corresponds to the $\bar{\mathrm{MS}}$--scheme), we obtain
\begin{align}
\int_0^1\mathrm{d}z\ \delta(1-z) \mathbf{I}_{\widetilde{a\imath}} \left( p_1,\hdots,p_m;\epsilon \right) =& -\frac{\alpha_s}{2\pi}\frac{1}{\Gamma(1-\epsilon)}  \sum_{\substack{b=1 \\ b \neq \widetilde{a\imath}}}^m \frac{\mathbf{T}_{\widetilde{a\imath}}\mathbf{T}_b}{\mathbf{T}_{\widetilde{a\imath}}^2} \left( \frac{4\pi\mu^2}{2\tilde{p}_{ai}p_b} \right)^\epsilon \mathcal{V}_{qg}(\epsilon),\label{eq:ana_I}\\
%\end{align}
%Note that $\mathbf{I}_{\widetilde{a\imath}}$ comes with a $\delta(1-z)$ prefactor and therefore $p_a = \tilde{p}_{ai}$ in this term.
%\begin{align}
\int_0^1 \mathrm{d}z\ \mathbf{P}_{\widetilde{a\imath}} \left( p_1,\hdots,p_m;z;\mu_F \right) =& \int_0^1 \mathrm{d}z\ \frac{\alpha_s}{2\pi}  \sum_{\substack{b=1 \\ b \neq \widetilde{a\imath}}}^m \frac{\mathbf{T}_{\widetilde{a\imath}}\mathbf{T}_b}{\mathbf{T}_{\widetilde{a\imath}}^2} \mathrm{ln} \frac{\mu_F^2}{2\tilde{p}_{ai}p_b} P_{qq}(z)= 0\;.
\end{align}
Note that $\mathbf{P}_{\widetilde{a\imath}}$ vanishes because $P_{qq}(z)$ is a pure plus distribution.
We finally obtain
\begin{align}
&\int_0^1 \mathrm{d}z\ \mathbf{H}_{\widetilde{a\imath}} \left( p_1,\hdots,p_a,\hdots,p_m;n;z \right)\nonumber\\ 
&= -\frac{\alpha_s}{2\pi} \sum_{\substack{b=1 \\ b \neq \widetilde{a\imath}}}^m \frac{\mathbf{T}_{\widetilde{a\imath}}\mathbf{T}_b}{\mathbf{T}_{\widetilde{a\imath}}^2} \int_0^1 \mathrm{d}z\ \left( \bar{K}^{qq}(z) + 2P_{qq}(z)\mathrm{ln}z + \tilde{K}^{qq}(z) + \mathcal{L}^{qq}(z;p_a,p_b,n)\right) \nonumber\\
%&= -\frac{\alpha_s}{2\pi} \sum_{\substack{b=1 \\ b \neq \widetilde{a\imath}}}^m \frac{\mathbf{T}_{\widetilde{a\imath}}\mathbf{T}_b}{\mathbf{T}_{\widetilde{a\imath}}^2} C_F\left\{ \frac{1}{2} + \frac{1}{2} - (5-\pi^2) +2\left( \frac{5}{4}-\frac{\pi^2}{3} \right) + \frac{7}{4} - \frac{\pi^2}{3} \right.\nonumber\\
%& + 2 \left[\mathrm{Li}_2 \left(1-\frac{1+v}{2}\frac{(\tilde{p}_{ai}+p_b)n}{\tilde{p}_{ai}n}\right)  + \mathrm{Li}_2 \left(1-\frac{1-v}{2}\frac{(\tilde{p}_{ai}+p_b)n}{\tilde{p}_{ai}n}\right) \right]\nonumber\\
%& +\left. \int_0^1 \mathrm{d}z\ (1+z)\mathrm{ln} \frac{n_{z}^2 \tilde{p}_{ai} p_b}{2 z (\tilde{p}_{ai} n_{z})^2} \right\}\\
&= -\frac{\alpha_s}{2\pi} \sum_{\substack{b=1 \\ b \neq \widetilde{a\imath}}}^m \mathbf{T}_{\widetilde{a\imath}}\mathbf{T}_b \left\{ \frac{1}{4} + 2 \left[\mathrm{Li}_2 \left(1-\frac{1+v}{2}\frac{(\tilde{p}_{ai}+p_b)n}{\tilde{p}_{ai}n}\right) + \mathrm{Li}_2 \left(1-\frac{1-v}{2}\frac{(\tilde{p}_{ai}+p_b)n}{\tilde{p}_{ai}n}\right) \right] \right. \nonumber\\
&\qquad +\left. \int_0^1 \mathrm{d}z\ (1+z)\mathrm{ln} \frac{n_{z}^2 \tilde{p}_{ai} p_b}{2 z (\tilde{p}_{ai} n_{z})^2} \right\}.\label{eq:ana_H}
\end{align}
The last line of Eq.~\eqref{eq:ana_H} is the only $z$-dependent contribution which cannot be integrated analytically.
Note in particular that $n_z$ implicitly depends on $z$ through Eq.~\eqref{eq:def_n}, where the momentum of the emitter
particle is given by $p_a = z\tilde{p}_{ai}$.

We remark that the introduction of the collinear counterterms in Eq.~\eqref{eq:IDcoll} is actually unnecessary,
since they give a vanishing contribution to the cross-section. This is due to $P_{qq}(z)$ being a pure plus-distribution
and the test-function with which it is convoluted not being $z$-dependent after we substituted $p_a = z \tilde{p}_{ai}$.
This is also true for differential cross-sections, since any partonic observable can be expressed without reference
to $z$.
The vanishing effect of collinear counterterms may also be understood from another point of view: As we do not actually
restrict the momenta of the ``identified'' partons, but integrate over them eventually, we have already collected
all singularities necessary to cancel those present in the virtual corrections. Hence, no collinear mass factorization
counterterms are required.

In summary, the integrated pseudo-dipoles for final-state singularities are given by 
\begin{align}
\int_{m+1}\mathrm{d}\sigma^\mathrm{S}\Big|_{\mathrm{FF}}^\mathrm{ID} = \sum_{\widetilde{a\imath}=1}^m \int_m \mathrm{d}\sigma^\mathrm{B}(p_1, \hdots, p_m)
 \left[\mathbf{I}_{\widetilde{a\imath}} + \int_0^1 \mathrm{d}z\  \mathbf{H}_{\widetilde{a\imath}} \right]_{\substack{\widetilde{a\imath} \in \left\{\mathrm{fin}\right\} \\ b \in \left\{\mathrm{fin}\right\}}}\;,
\end{align}
where the $\mathbf{I}$- and $\mathbf{H}$-operator are given in the Eq.~\eqref{eq:ana_I} and Eq.~\eqref{eq:ana_H} respectively.
Pseudo-dipoles describing final-state singularities with final-state and initial-state spectators are identical, apart from the replacement of $b$ being the initial-state spectator instead of a final-state one. Hence we obtain the same formulae for $\left.\int_{m+1} \mathrm{d}\sigma^\mathrm{S} \right|_{\mathrm{FI}}^\mathrm{ID}$.
In contrast to standard FI-dipoles, we do not obtain a $\mathbf{K}$-operator at this stage. 
For completeness, we quote the integrated standard FI-dipoles ($\mathcal{D}_{ij}^{a}$) as those contribute to the difference when changing the implementation from standard to pseudo-dipoles.
Labeling the emitter parton $j$ and the initial-state colour spectator $b$, its integral reads
\begin{align}
\int_{m+1} \mathrm{d}\sigma^\mathrm{S} \Big|_{\mathrm{FI}}^\mathrm{CS} &= \mathcal{N}_{in}\int_{m+1} \sum_{\left\{m+1\right\}} \mathrm{d}\phi_{m+1} \frac{1}{S_{\left\{m+1\right\}}} \sum_{\substack{\mathrm{pairs} \\ j,i}} \sum_{b} \mathcal{D}_{ij}^{b}F^{(m)}_J \bigg|_{\substack{j \in \left\{\mathrm{fin}\right\} \\ b \in \left\{\mathrm{in}\right\}}}\\
&= \sum_{\widetilde{\imath\jmath}=1}^m  \int_m 
\left[ \mathrm{d}\sigma^\mathrm{B}(p_a;p_1, \hdots, p_m)\mathbf{I}_{\widetilde{\imath\jmath}} + \int_0^1 \mathrm{d}x\ \mathrm{d}\sigma^\mathrm{B}(xp_b;p_1, \hdots, p_m) \mathbf{K}_{b}^\mathrm{CS-FI} \right]_{\substack{\widetilde{\imath\jmath} \in \left\{\mathrm{fin}\right\} \\ b \in \left\{\mathrm{in}\right\}}},
\end{align}
where the sum over $j$ ($\widetilde{\imath\jmath}$) extend over all non-identified final-state partons and the sum over $b$ extends over all initial-state spectators. $\mathbf{I}_{\widetilde{\imath\jmath}}$ is given in Eq.~\eqref{eq:ana_I} (with $\widetilde{a\imath} \to \widetilde{\imath\jmath}$) and 
\begin{align}
\mathbf{K}_{b}^\mathrm{CS-FI}(x) = -\frac{\alpha_s}{2\pi} \sum_{b} \mathbf{T}_{\widetilde{\imath\jmath}} \mathbf{T}_b & \left\{ \left( \frac{2}{1-x}\mathrm{ln}\frac{1}{1-x} \right)_+ + \frac{2}{1-x}\mathrm{ln}(2-x)\right.\nonumber\\
&\left. -\frac{\gamma_{\widetilde{\imath\jmath}}}{\mathbf{T}_{\widetilde{\imath\jmath}}^2} \left[ \left(\frac{1}{1-x}\right)_+ + \delta(1-x) \right] \right\}
\end{align}
with $\gamma_{\widetilde{\imath\jmath}}$ the collinear anomalous dimensions given for example in appendix~C of Ref.~\cite{Catani:1996vz}.

\subsubsection{Initial-state singularities: integrated form}
In the following we will quote results of the integrated pseudo-dipoles for initial-state singularities.
We follow the labelling in Fig.~\ref{fig:in_sing_assignment}, i.e. parton $a$ is the initial-state emitter and $b$ the final-state spectator. As for standard dipoles we have to sum over all possible internal parton flavors $\widetilde{a\imath}$ which can be obtained in the branching $a \to \widetilde{a\imath}+i$, and which lead to a non-vanishing Born matrix element. Including the collinear mass factorization counterterms, we obtain the following result:
\begin{align}
&\int_{m+1} \mathrm{d}\sigma^\mathrm{S} \Big|_{\mathrm{IF}}^\mathrm{ID} = \mathcal{N}_{in} \frac{1}{n_s(a)\Phi(p_a)} \int_{m+1} \sum_{\left\{m+1\right\}} \mathrm{d}\phi_{m+1} \frac{1}{S_{\left\{m+1\right\}}} \sum_{\substack{\mathrm{pairs} \\ a,i}} \sum_{b} \mathcal{D}_{b}^{(n)ai}F^{(m)}_J \bigg|_{\substack{a \in \left\{\mathrm{in}\right\} \\ b \in \left\{\mathrm{fin}\right\}}},\\
&\int_{m+1} \mathrm{d}\sigma^\mathrm{S} \Big|_{\mathrm{IF}}^\mathrm{ID} + \int_{m} \mathrm{d}\sigma^\mathrm{C} = \sum_{\widetilde{a\imath}} \int_m 
\bigg[ \mathrm{d}\sigma^\mathrm{B}(p_a, p_1, \hdots, p_m) \delta_{a \widetilde{a\imath}} \mathbf{I}_{\widetilde{a\imath}}\nonumber\\
&\hspace*{3cm} + \int_0^1 \mathrm{d}x\  \mathrm{d}\sigma^\mathrm{B}(xp_a, p_1, \hdots, p_m) \left( \mathbf{K}^{\mathrm{ID-IF}}_{a, \widetilde{a\imath}}+ \mathbf{P}^{\mathrm{IF}}_{a, \widetilde{a\imath}} \right) \bigg]_{\substack{\widetilde{a\imath} \in \left\{\mathrm{in}\right\} \\ \tilde{b} \in \left\{\mathrm{fin}\right\}}},\label{eq:integratedIFpseudodipole}
\end{align}
Next to the sum over all possible $(a,i)$-pairs the sum over $b$ includes all identified final-state colour spectators.
Note that in Eq.~\eqref{eq:integratedIFpseudodipole} we refer to the color spectator as $\tilde{b}$ rather than $b$.
This is due to our identification scheme for IF-pseudo-dipoles described in sec.~\ref{sec:exploitPseudoDipoles}, where the momentum of the colour spectator is mapped from $p_b \to \tilde{p}_b$.
Due to this scheme which is opposed to the original use case in Ref.~\cite{Catani:1996vz}, we are required to substitute $p_b$ with $\tilde{p}_b$ in the differential form of the IF-pseudo-dipoles as well.
Apart from the correlated Born matrix-element, the momentum of parton $b$ enters only in the kinematic variable $v_{i,ab}$
(see Eq.~\eqref{eq:v_iab}). We can adjust for this by making the replacement
\begin{align}
v_{i,ab}\to\tilde{v}_{i,ab} = \frac{p_a\tilde{p}_b}{p_i(p_a+\tilde{p}_b)}
\end{align}
in the splitting functions in Eq.~\eqref{eq:VIDinitial}.
The soft and collinear limit of the splitting function are not affected by this change.

The insertion operator $\mathbf{I}_{\widetilde{a\imath}}$ in Eq.~\eqref{eq:integratedIFpseudodipole}
is obtained by replacing $b\to\bar{b}$ in Eq.~\eqref{eq:ana_I} and using
\begin{align}
\mathbf{K}^{\mathrm{ID-IF}}_{a, \widetilde{a\imath}}(p_1, \hdots, p_m; p_a, n, x) &= -\frac{\alpha_s}{2\pi} \left\{ \sum_{\tilde{b}=1}^m \frac{\mathbf{T}_{\tilde{b}}\mathbf{T}_{\widetilde{a\imath}}}{\mathbf{T}_{\widetilde{a\imath}}^2} \left[ \overline{K}^{a,\widetilde{a\imath}}(x) + \tilde{K}^{a,\widetilde{a\imath}}(x) + \mathcal{L}^{a,\widetilde{a\imath}}(x;p_a,\tilde{p}_b,n) \right] \right.\nonumber\\
& \left. + K_{\mathrm{F.S.}}^{a,\widetilde{a\imath}}(x) \right\}.
\end{align}
The $\mathbf{P}^{\mathrm{IF}}_{a, \widetilde{a\imath}}(p_1, \hdots, p_m; xp_a, x, \mu_F^2)$-operator is the same as for standard dipoles and given in Eq.~(8.39) of Ref.~\cite{Catani:1996vz}.
This result differs from the one using standard dipoles only by the $\mathbf{K}$-operator, which is given by
\begin{align}
\mathbf{K}^{\mathrm{CS-IF}}_{a, \widetilde{a\imath}}(x) = -\frac{\alpha_s}{2\pi} & \left\{ \sum_{k=1}^m \frac{\mathbf{T}_k\mathbf{T}_{\widetilde{a\imath}}}{\mathbf{T}_{\widetilde{a\imath}}^2} \left[ \overline{K}^{a,\widetilde{a\imath}}(x) -\delta^{a\ \widetilde{a\imath}} \mathbf{T}_a^2\left[ \left( \frac{2}{1-x}\mathrm{ln}\frac{1}{1-x} \right)_+ + \frac{2}{1-x}\mathrm{ln}(2-x) \right] \right] \right.\nonumber\\
& \left.+ K_{\mathrm{F.S.}}^{a,\widetilde{a\imath}}(x) \right\}.
\end{align}
For the case of initial-state spectators there is no difference to the standard dipole.

\section{Application to $W^+W^-b\bar{b}$ Production }\label{sec:ee_bwbw}
We have tested the above described resonance-aware subtraction by means of pseudo-dipoles
in two reactions: $e^+e^- \to W^+W^-b\bar{b}$ and $pp \to W^+W^-j_bj_b$. In the following we are going to compare results
obtained with standard CS dipoles to those obtained with pseudo-dipoles for fixed NLO QCD predictions.
In Sec.~\ref{sec:sing_limits}, we first examine the cancellation of divergences between the
real-emission matrix elements and the different dipoles using ensembles of trajectories in phase-space,
which approach the collinear and soft limits in a controlled way. Following this, we compare physical
cross-sections calculated with the different subtraction techniques while paying special attention
to the rate of convergence in the Monte-Carlo integration.

\subsection{Singular limits}\label{sec:sing_limits}
In this section we validate the implementation of the differential form of the dipole insertion
operators in Eqs.~\eqref{eq:pseudo_dipole_fs} and~\eqref{eq:pseudo_dipole_is} by testing the behavior
of the subtracted real-emission corrections in their singular limits.

\subsubsection{$e^+e^- \to W^+W^-b\bar{b}$}
The sole real-emission correction to the process $e^+e^- \to W^+W^-b\bar{b}$ at NLO is the process
with an additional gluon in the final state. It develops singularities when the gluon
is soft, and when it is collinear to the $b$ (anti-)quark. 
%We choose the labelling of momenta as in \texttt{Sherpa}
%\begin{align}
%e^-(p_0) + e^+(p_1) \to W^+(p_2) + W^-(p_3) + j(p_4) + b(p_5)+\bar{b}(p_6)
%\end{align}
To parametrize these limits we use the scaled virtualities,
\begin{align}
y_{gb} = \frac{2p_gp_b}{\left(p_g+p_b+p_{\mathrm{b}}\right)^2} 
\hspace{1cm}\mathrm{and}\hspace{1cm}
y_{g\bar{b}} = \frac{2p_gp_{\bar{b}}}{\left(p_g+p_b+p_{\mathrm{b}}\right)^2}\;.
\end{align}
In the collinear regions, $y_{gb}\to0$ or $y_{g\bar{b}}\to0$,
while in the soft limit $y_{gb}y_{g\bar{b}}\to0$.
\begin{figure}[t]
\begin{center}
\begin{minipage}[t]{0.49\textwidth}
\includegraphics[width=0.97\textwidth]{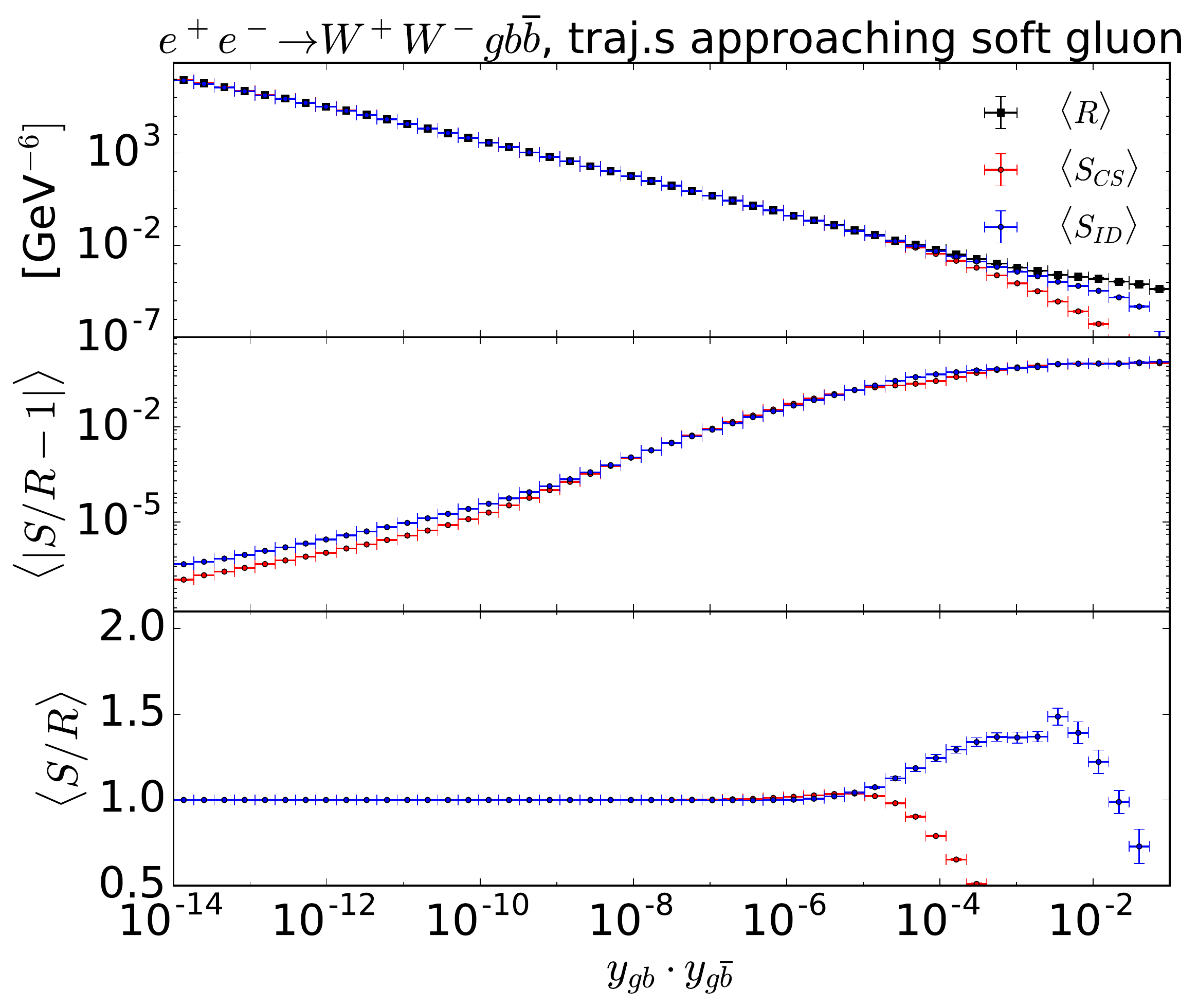}
\end{minipage}
\hfill
\begin{minipage}[t]{0.49\textwidth}
\includegraphics[width=\textwidth]{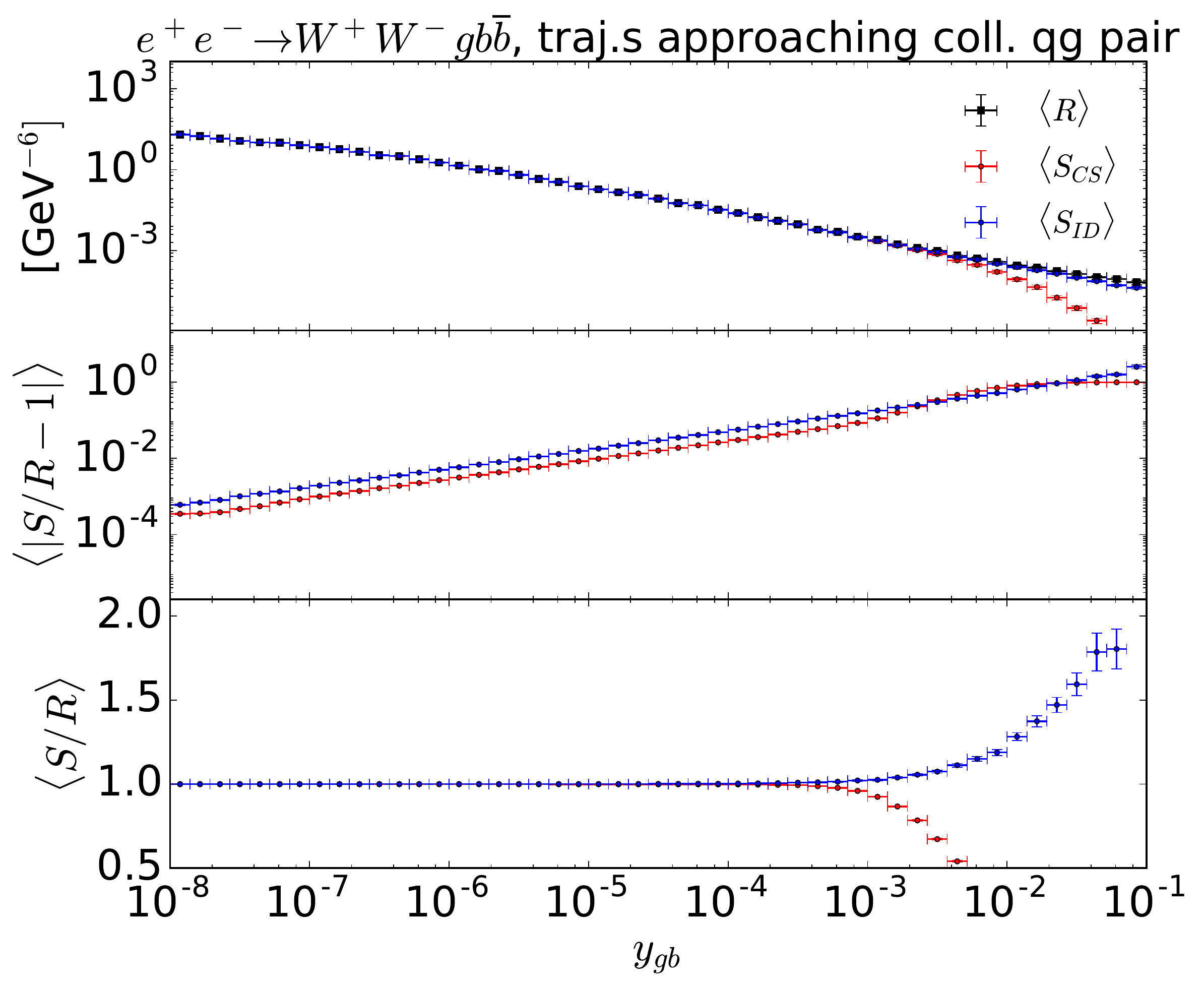}
\end{minipage}
\caption{Matrix elements $R=\left|\mathcal{M}_R\right|^2$ and sum of associated dipoles $S=\sum \mathcal{D}$ (standard CS-dipoles in red, pseudo-dipoles in blue) for the process $e^+e^- \to W^+W^-gb\bar{b}$. Both dipole-types are evaluated on the same trajectory like described in the text.\newline
Right: Average over trajectories in phase-space with increasingly soft gluon, left: Average over trajectories with increasingly collinear qg-pair.}\label{fig:eewwgbb}
\end{center}
\end{figure}
We construct ensembles of phase-space points that we refer to as phase-space trajectories
as follows: random phase-space points are sampled according to the real-emission cross section
in the narrow-width approximation, while requesting three identified jets according to the
Durham jet algorithm~\cite{Catani:1991hj}. For each phase-space point, the kinematical configuration
is scaled according to the algorithm described in App.~\ref{sec:momentum_scaling} in order to
obtain a sequence of points that approach the soft or collinear limit.

Figure~\ref{fig:eewwgbb} shows the values of the real-emission matrix-element,
$R=\left|\mathcal{M}_R\right|^2$ and the sum of the associated dipoles ($S=\sum \mathcal{D}$),
as well as their ratio for the soft limit (left) and
the $bg$--collinear limit (right) limit. 
We have averaged -- for all sub-plots -- over all entries within a bin.
In doing so, the ratio plots enable us to assess the pointwise convergence of $S/R$ best.
It can be seen that the cancellation of divergences
occurs in both the standard CS subtraction method and in the pseudo-dipole approach,
but that the pseudo-dipoles converge faster towards the real-emission matrix element,
which can be seen in the lower panels of the figures. 

\subsubsection{$pp \to W^+W^-j_bj_b$}
Next we consider the analogue to the above example at hadron colliders, namely the reaction
$pp \to W^+W^-j_bj_b$, where $j_b$ indicates a $b$-tagged jet. To parametrize the soft and
collinear trajectories in this process, we use the following variables
\begin{align}
v_i^0 = \frac{p_ip_0}{p_0p_1}
\hspace{1cm}\mathrm{and}\hspace{1cm}
v_i^1 = \frac{p_ip_1}{p_0p_1}
\end{align}
where $p_0$ and $p_1$ are the momenta of the initial-state partons and $p_i$ is the momentum
of the additional parton in the real correction. The phase-space trajectories are again
constructed by generating random phase-space points with well separated partons, which are
then modified according to the algorithm in App.~\ref{sec:momentum_scaling} to obtain
additional phase-space points that approach the singular limits.

\begin{figure}[t]
\begin{center}
\begin{minipage}[t]{0.49\textwidth}
\includegraphics[width=\textwidth]{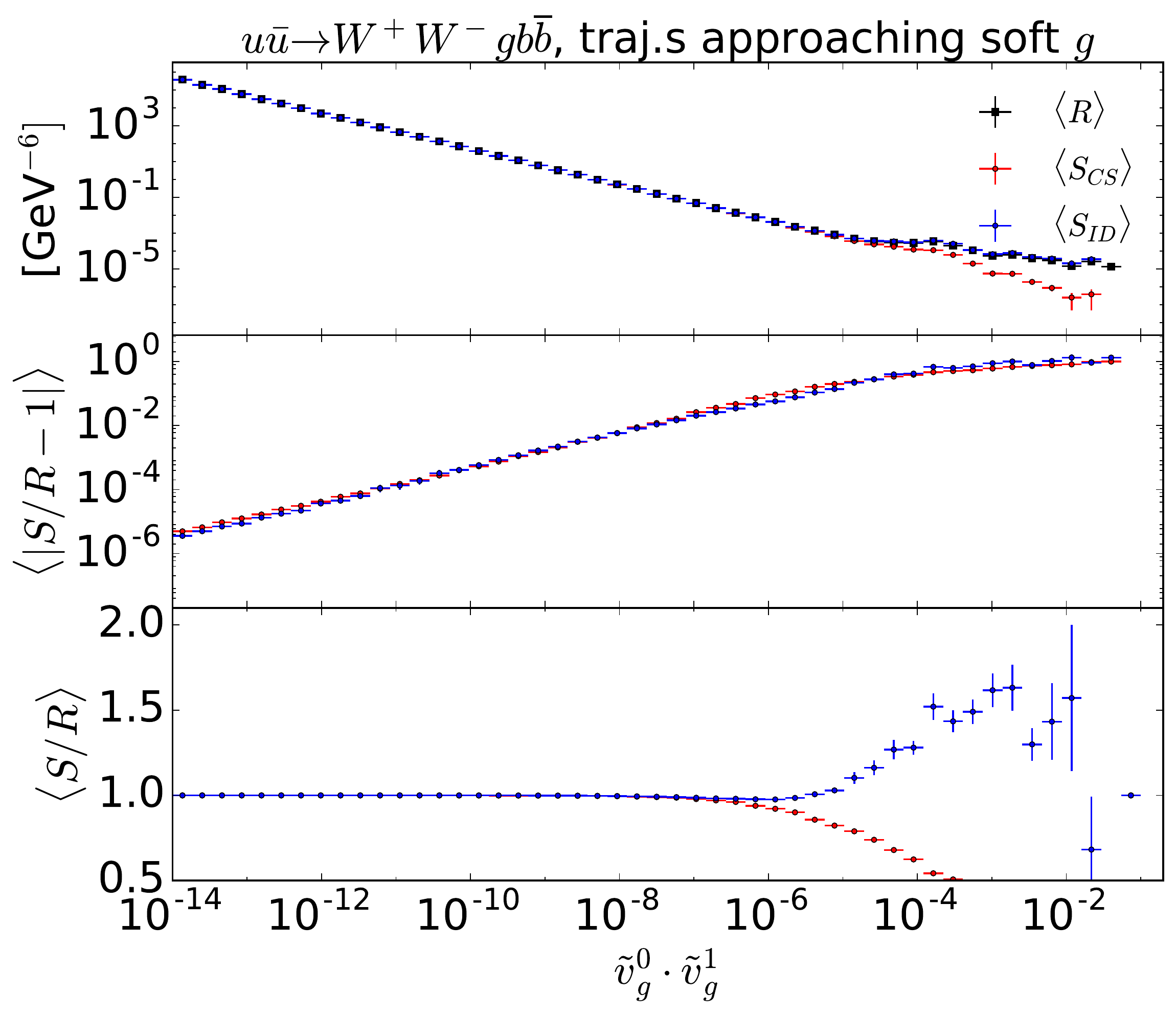}
\end{minipage}
\hfill
\begin{minipage}[t]{0.49\textwidth}
\includegraphics[width=\textwidth]{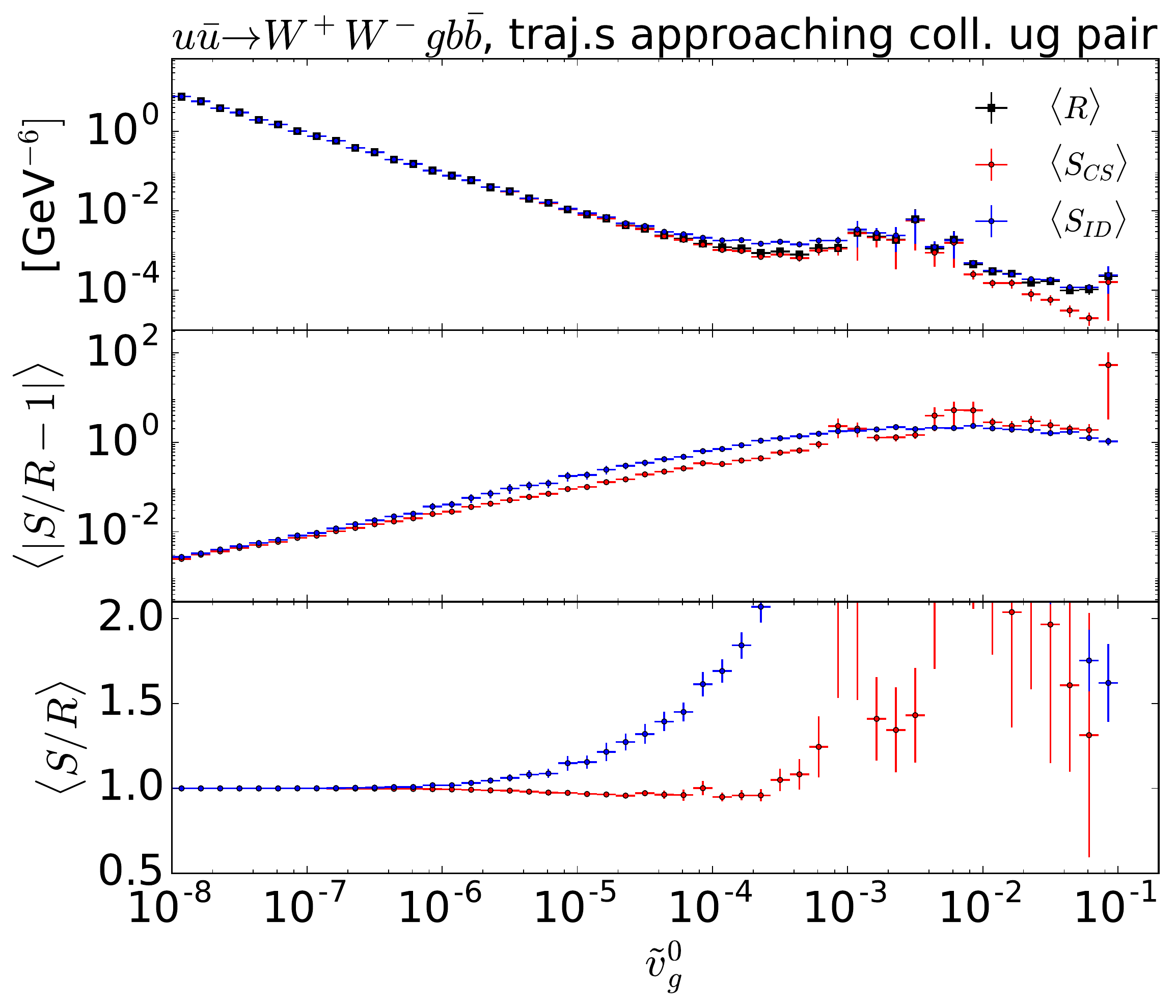}
\end{minipage}
\begin{minipage}[t]{0.49\textwidth}
\includegraphics[width=\textwidth]{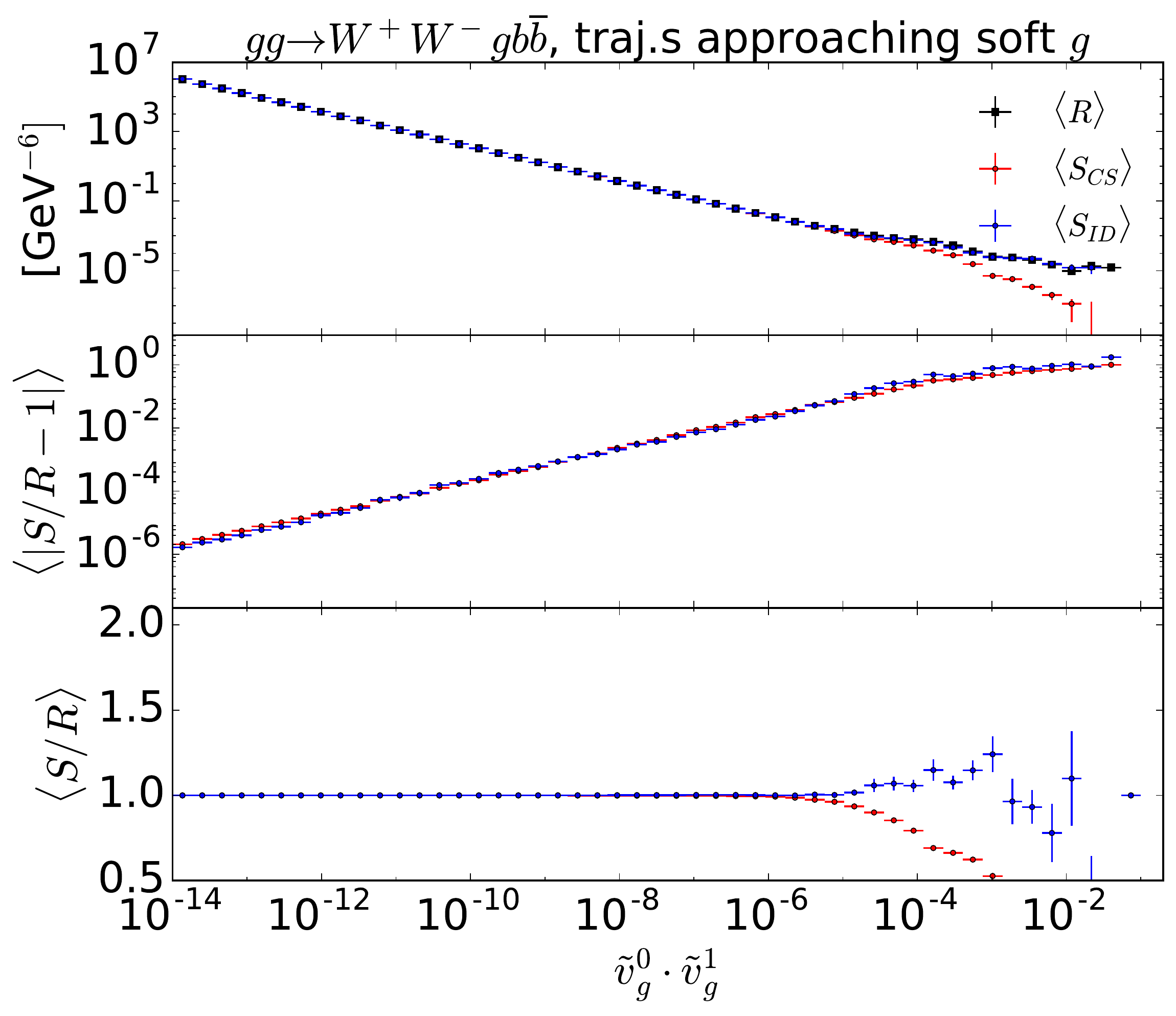}
\end{minipage}
\hfill
\begin{minipage}[t]{0.49\textwidth}
\includegraphics[width=\textwidth]{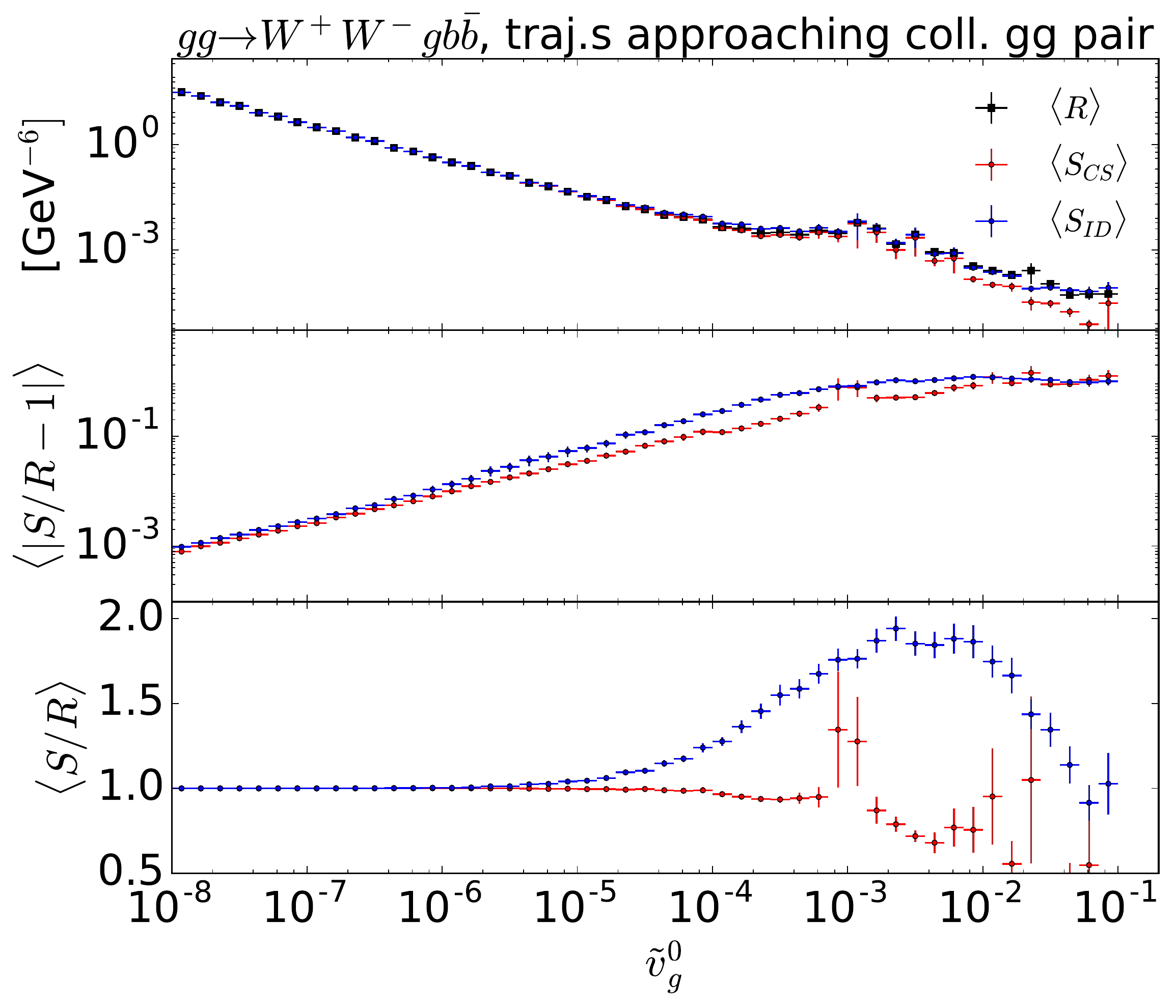}
\end{minipage}
\caption{Matrix elements $R=\left|\mathcal{M}_R\right|^2$ and sum of associated dipoles $S=\sum \mathcal{D}$ (standard CS-dipoles in red, pseudo-dipoles in blue) for the processes $u\bar{u} \to W^+W^-gb\bar{b}$ (top) and $gg \to W^+W^-gb\bar{b}$ (bottom). Both dipole-types are evaluated on the same trajectories like described in the text.\newline
Left: Average over trajectories in phase-space with increasingly soft gluon. Right: Average over trajectories with increasingly collinear partons.}\label{fig:ppwwgbb1}
\end{center}
\end{figure}

\begin{figure}[t]
\begin{center}
\begin{minipage}[t]{0.49\textwidth}
\includegraphics[width=\textwidth]{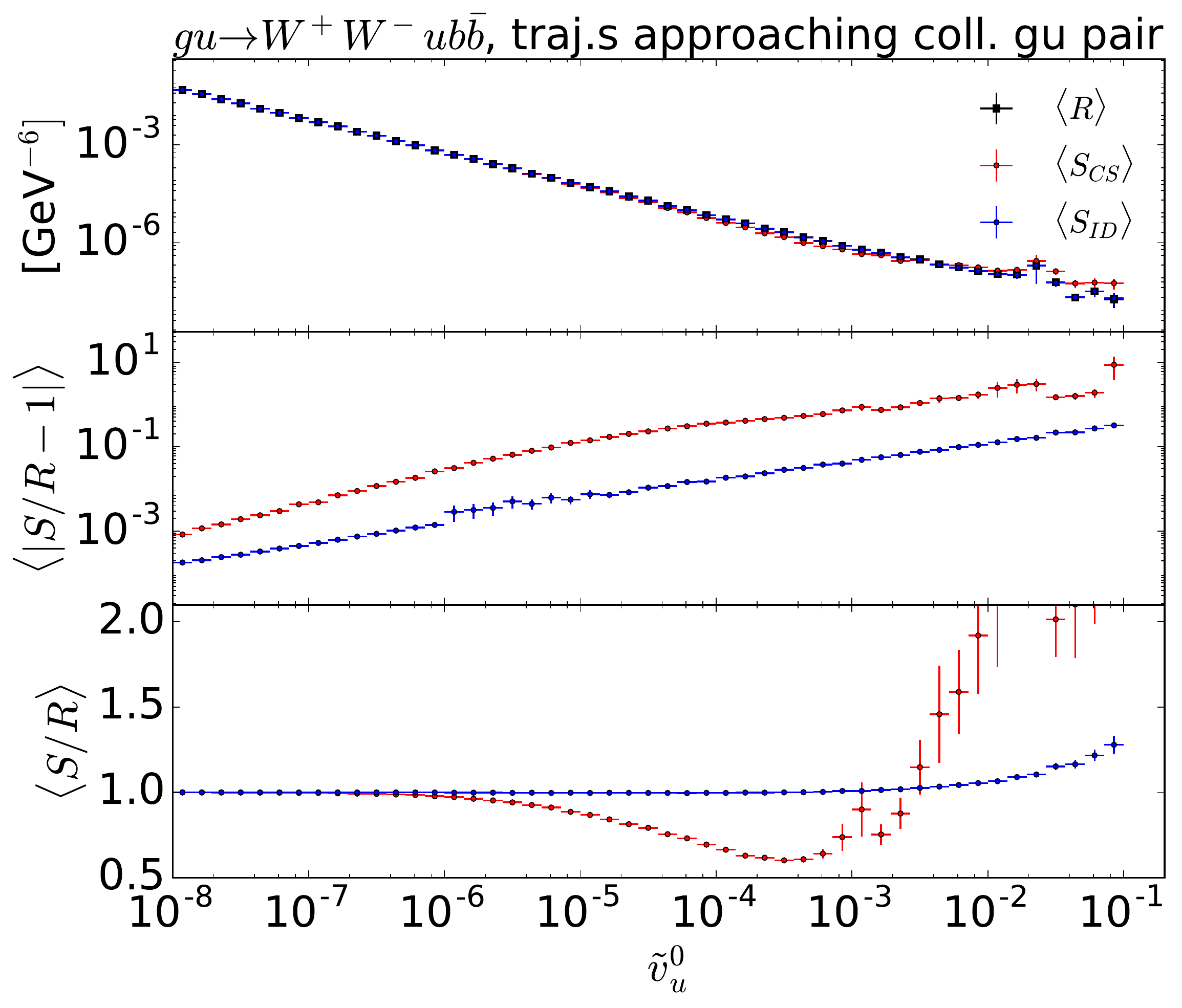}
\end{minipage}
\hfill
\begin{minipage}[t]{0.49\textwidth}
\includegraphics[width=\textwidth]{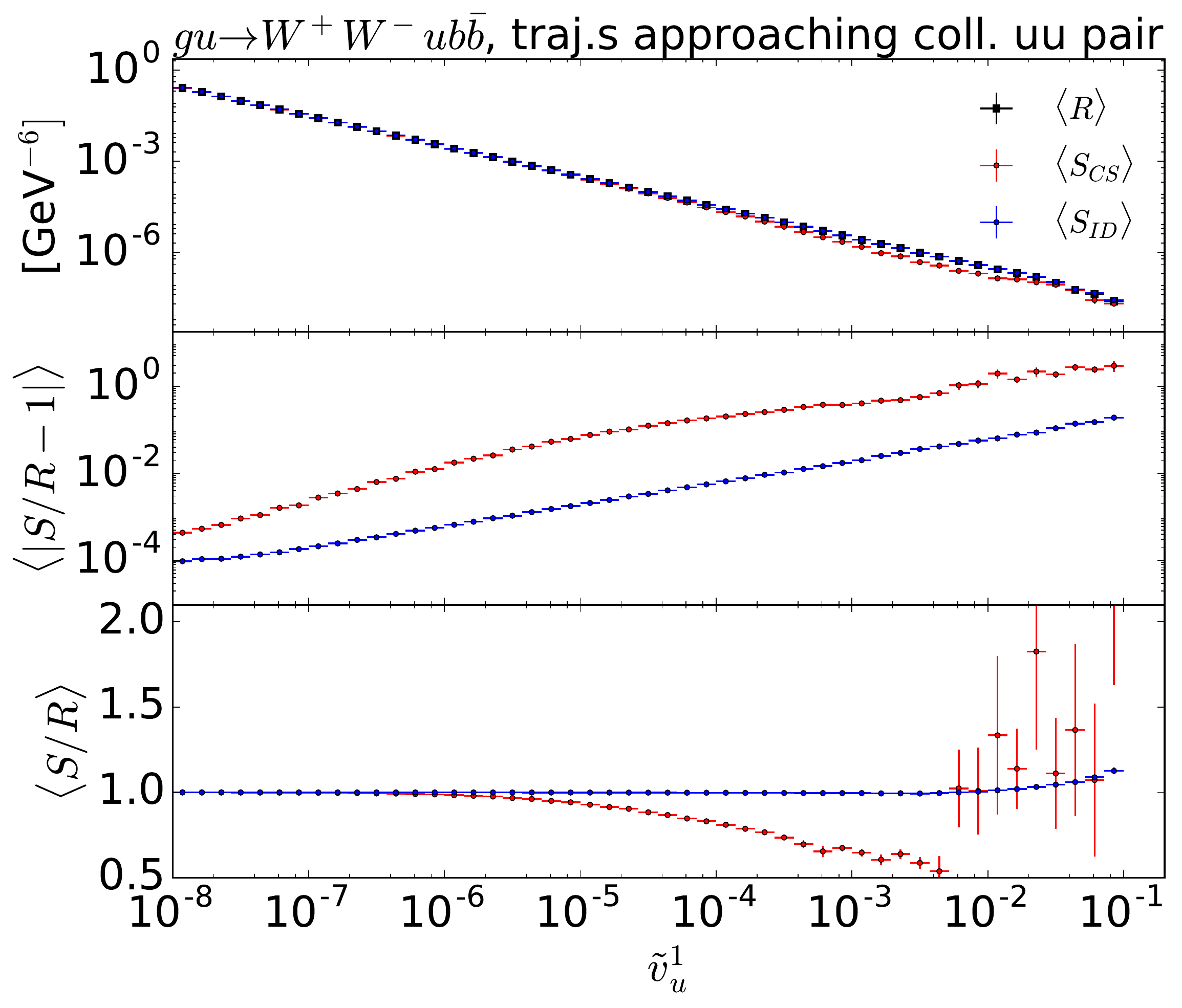}
\end{minipage}
\caption{Matrix elements $R=\left|\mathcal{M}_R\right|^2$ and sum of associated dipoles $S=\sum \mathcal{D}$ (standard CS-dipoles in red, pseudo-dipoles in blue) for the process $ug \to W^+W^-ub\bar{b}$.
Both dipole-types are evaluated on the same trajectories, with increasingly collinear partons, like described in the text.\newline
Left: The $u$-quark in the final state becomes collinear to the first initial-state parton.
Right: The $u$-quark in the final state becomes collinear to the second initial-state parton.}\label{fig:ppwwgbb2}
\end{center}
\end{figure}

Figures~\ref{fig:ppwwgbb1} and~\ref{fig:ppwwgbb2} display the average values of the real-emission
matrix element and the corresponding dipole subtraction terms, as well as their ratio -- averaged within each bin -- for processes with an additional gluon (Fig.~\ref{fig:ppwwgbb1}) and an additional
quark (Fig.~\ref{fig:ppwwgbb2}) in the final state. The former develop both soft and collinear singularities,
shown in the left and right panels of Fig.~\ref{fig:ppwwgbb1}, respectively, while the latter
only feature collinear singularities. Combining all tests, we validate all four initial-state
splitting functions in Eq.~\eqref{eq:VIDinitial}.

It can be seen that both the standard CS- as well as the pseudo-dipoles converge towards the real correction.
In the soft limit, which is the numerically more important one, the pseudo-dipoles converge faster.
This is also the case for the collinear limit probed in Fig.~\ref{fig:ppwwgbb2}.
However, for the collinear limits tested on the right-hand side of Fig.~\ref{fig:ppwwgbb1} this is not the case.
As a reason for this we suspect that the pseudo-dipole splitting functions Eq.~\eqref{eq:VIDinitial}
converge more slowly towards the associated Altarelli-Parisi functions than the standard CS ones.
%\todo{Seb: suspect splitting functions, since recoil scheme is by construction better in ID-case.}

\subsection{Physical cross sections}
In this section we present first results validating the pseudo-dipole subtraction method for resonance-aware processes
at the level of observable cross sections and distributions.
Results are cross-checked using two different implementations of our new algorithm within the public event generation framework
Sherpa~\cite{Gleisberg:2003xi,Gleisberg:2008ta}, one using the matrix-element generator
\texttt{AMEGIC++}~\cite{Krauss:2001iv,Gleisberg:2007md}, and one using the new interface between
\texttt{Sherpa} and~\texttt{OpenLoops}~\cite{Jones:2017giv}. In this interface, the color-correlated
Born matrix-elements are imported from \texttt{OpenLoops} libraries~\cite{Cascioli:2011va},
while the splitting function is calculated in \texttt{Sherpa}
and the integration is performed using the techniques implemented in \texttt{AMEGIC++}~\cite{Krauss:2001iv}.

\subsubsection{$e^+e^- \to W^+W^- b\bar{b}$}
Again we investigate first the reaction $e^+e^- \to W^+W^-b\bar{b}$.
We vary the center-of-mass energy of the collider to obtain predictions below, at and above the top-quark pair
production threshold, and we do not include the effects of initial-state radiation. We require two hard jets
at $y=(5\textrm{ GeV}/E_{\rm cms})^2$ defined according to the Durham jet algorithm~\cite{Catani:1991hj}.
The running of the strong coupling is evaluated at two loops, and the reference value is set to
$\alpha_s(M_Z^2)=0.118$, where $M_Z=91.1876~{\rm GeV}$.

\begin{table}[t]
\begin{center}
  \begin{tabular}{|c|c|c|c|c|c|c|}\hline
    &\multicolumn{2}{c|}{$\sqrt{s}=3m_W$}&\multicolumn{2}{c|}{$\sqrt{s}=2m_t$}
    &\multicolumn{2}{c|}{$\sqrt{s}=4m_t$} \\
    \cline{2-7}
$\sigma\mathrm{[fb]}$ & CS & ID & CS & ID & CS & ID \\
\hline
RS & $-0.00772(6)$ & $-0.00140(5)$ & $-0.52(3)$ & $-2.85(1)$ & $-9.5(4)$ & $-5.3(1)$ \\
BVI & $0.16143(13)$ & $0.15506(13)$ & $148.07(9)$ & $150.55(9)$ & $230.0(2)$ & $226.0(2)$ \\
$\sum$ & $0.15371(14)$ & $0.15366(14)$ & $147.55(9)$ & $147.70(9)$ & $220.5(4)$ & $220.7(2)$ \\\hline
\end{tabular} 
\caption{NLO cross sections for $e^+e^- \to W^+W^-b\bar{b}$ at $\mu_R = m_t$
  and varying center-of-mass energy, computed using standard CS subtraction terms (CS)
  or pseudo-dipoles (ID). The subtracted real-emission contributions (RS) were calculated
  using $10^7$ phase-space points. The Born, virtual corrections and integrated subtraction
  terms (BVI) were calculated using $3\cdot10^6$ phase-space-points.}\label{tab:RSBVI}
\end{center}
\end{table} 

\begin{figure}[tbp]
\begin{center}
\includegraphics[width=0.625\textwidth]{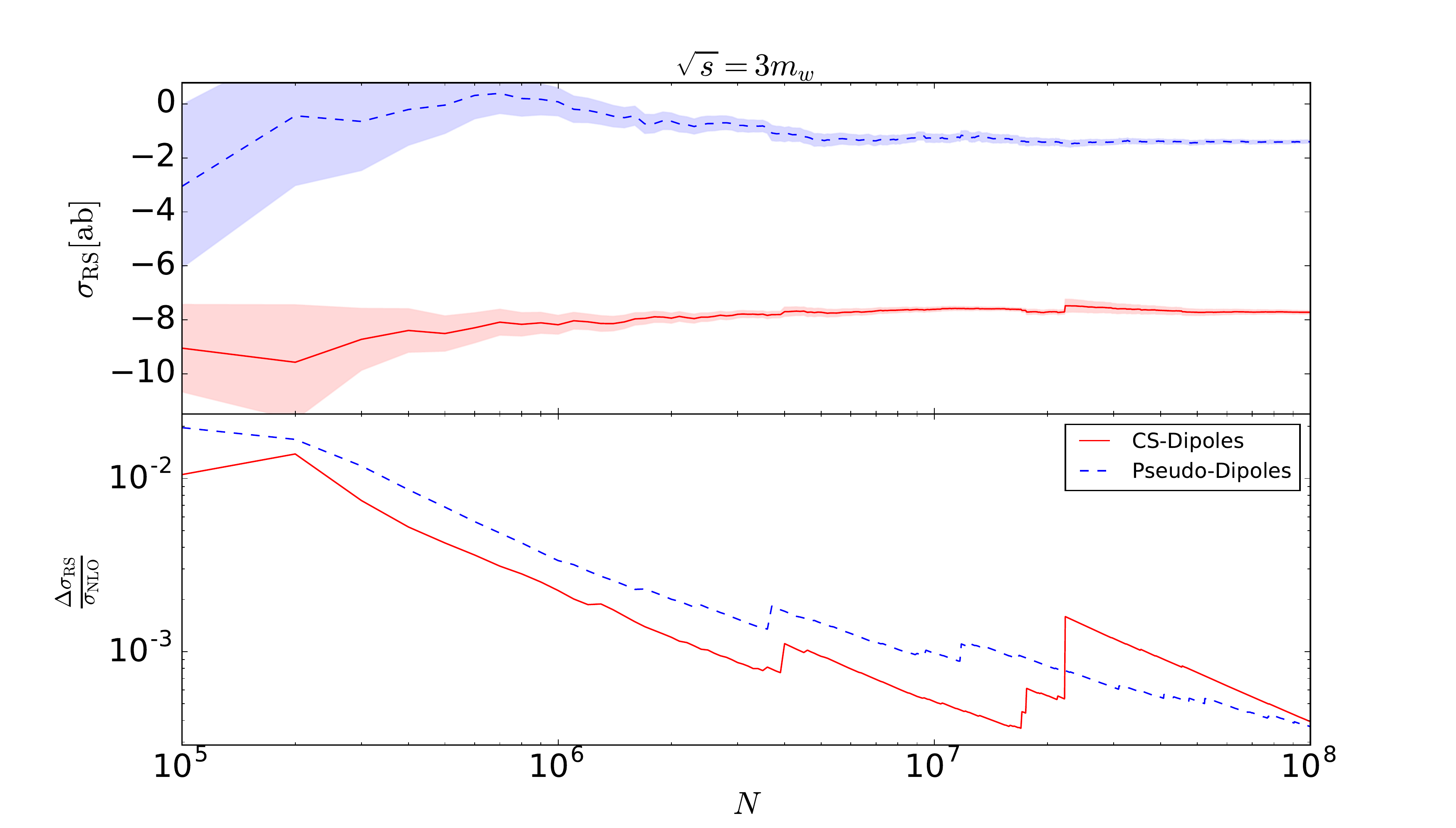}\\
\includegraphics[width=0.625\textwidth]{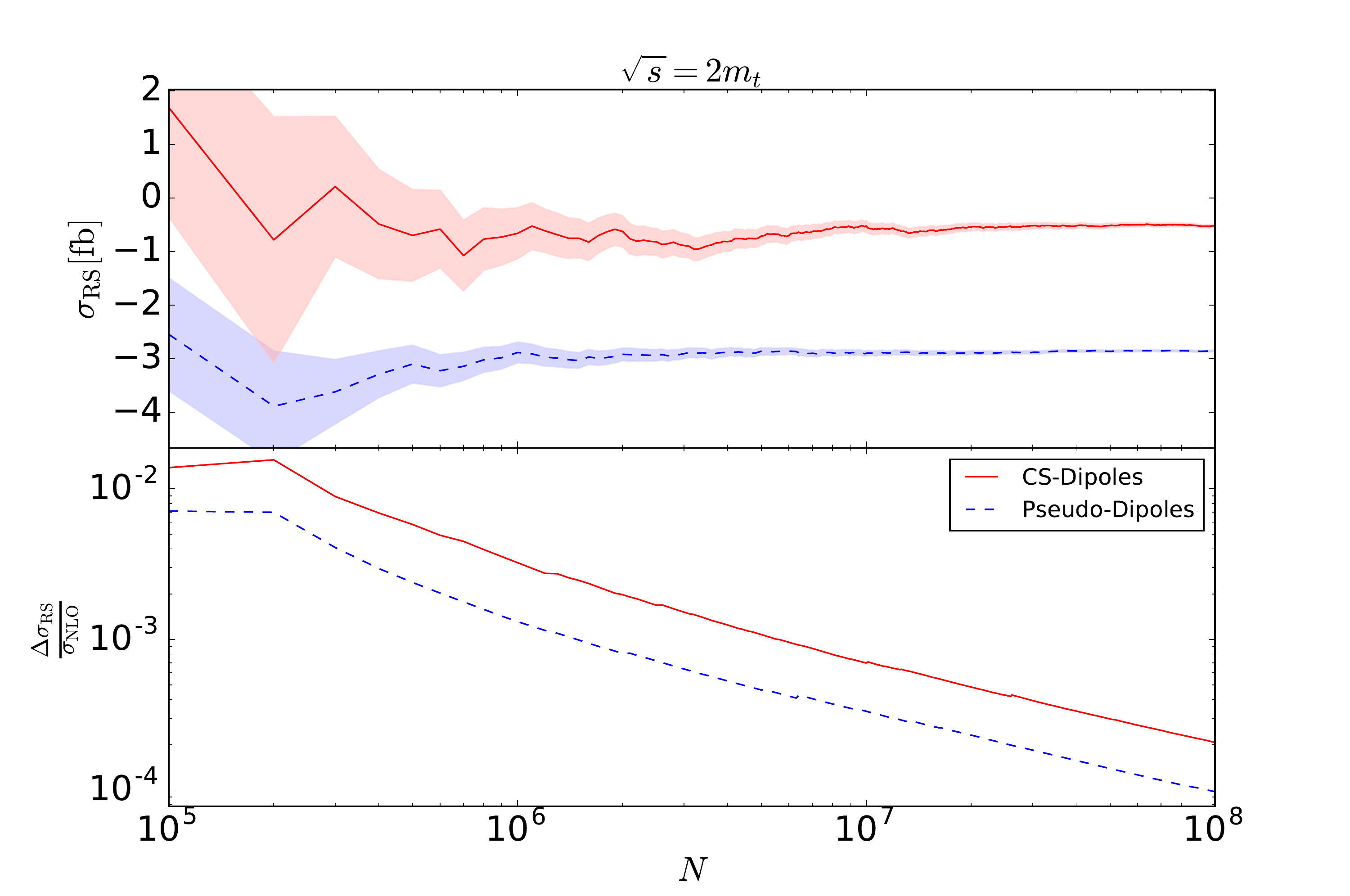}\\
\includegraphics[width=0.625\textwidth]{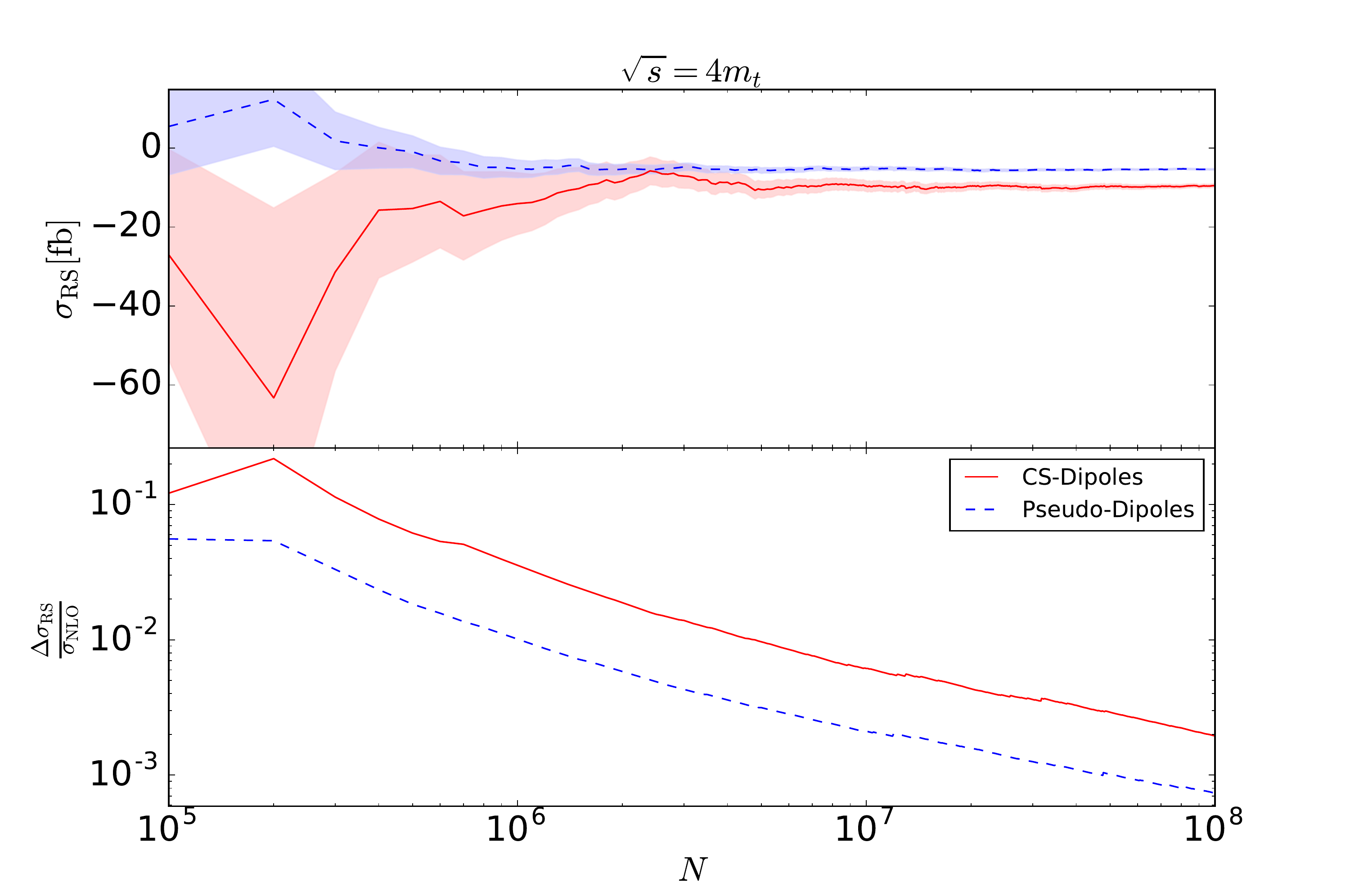}
\caption{Evolution of the Monte-Carlo integration results for the subtracted real-emission
  contribution to the total cross section in $e^+e^- \to W^+W^-b\bar{b}$ at varying
  center-of-mass energy over number of sampled phase-space points $N$. From top to bottom: $\sqrt{s} = 3m_W$, $\sqrt{s} = 2m_t$
  and $\sqrt{s} = 4m_t$. Red solid lines show results from standard CS-dipoles, while
  blue dashed lines correspond to pseudo-dipoles. The colored bands in the upper panels
  and the lines in the lower panels show the one $\sigma$ statistical uncertainty
  of the Monte-Carlo integration.}\label{fig:stat_ee_bwbw}
\end{center}
\end{figure}

Table~\ref{tab:RSBVI} shows the total cross sections as well as the individual contributions
from the subtracted real-emission terms (RS) as well as Born, virtual corrections and
integrated subtraction terms (BVI). As expected, the RS and BVI contributions differ
between the standard CS subtraction method and the pseudo-dipole approach, but their sum
agrees within the statistical accuracy of the Monte-Carlo integration.
The cross section is significantly enhanced at and above the production threshold
for a top-quark pair.
For those two center-of-mass energies, we expect the pseudo-dipoles to give a more physical
interpretation of the subtraction term and thus a reduced variance during the integration.
This is confirmed in Fig.\ref{fig:stat_ee_bwbw}, which shows the evolution
of the Monte-Carlo error during the integration.
In the case of pseudo-dipole subtraction at or above threshold, the uncertainty is indeed
substantially lower than for standard CS-dipoles. Below threshold the performance of
pseudo-dipole subtraction is similar to the standard technique.

\begin{figure}[tbp]
\begin{center}
\includegraphics[width=0.49\textwidth]{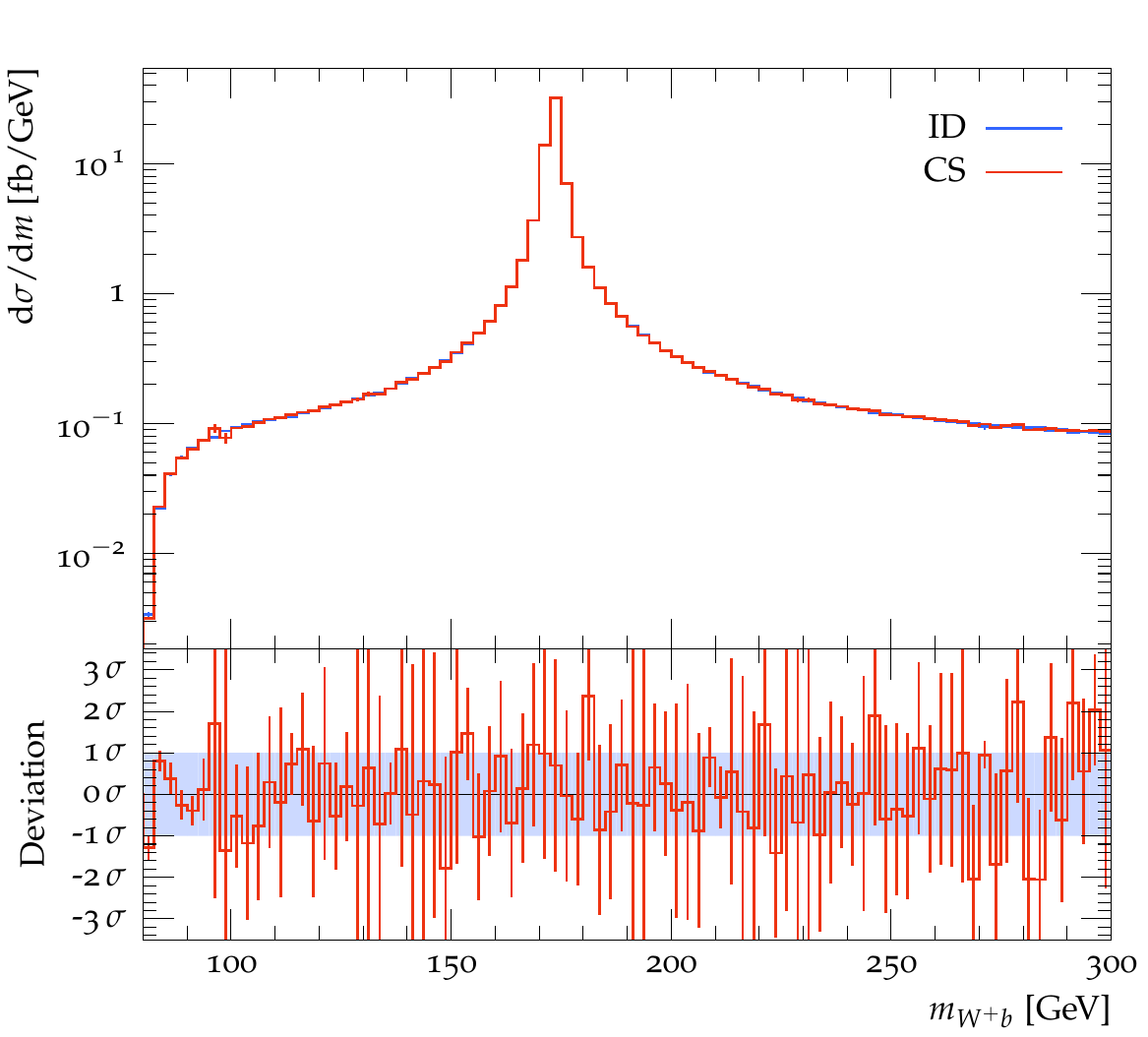}\hfill %
\includegraphics[width=0.49\textwidth]{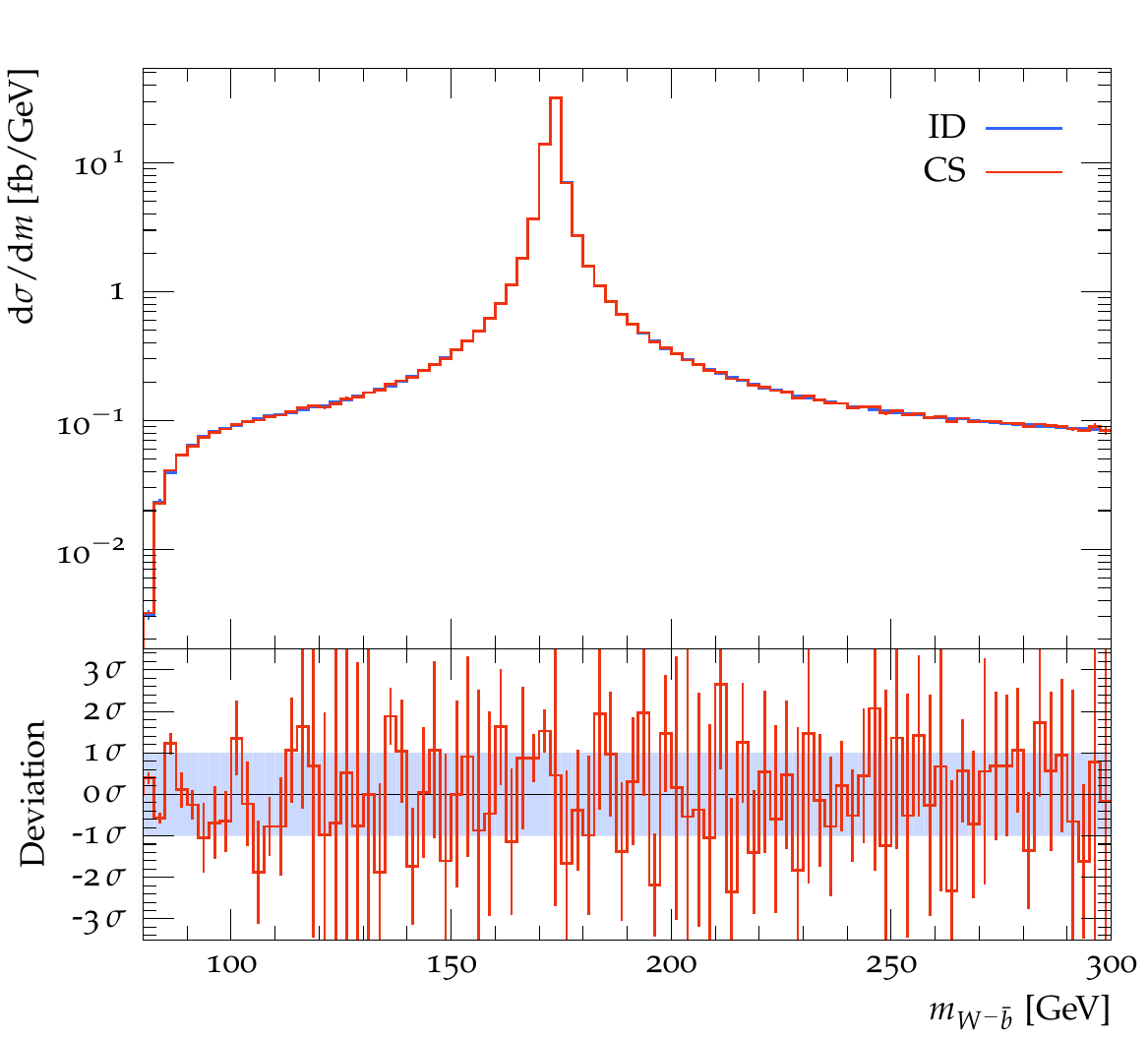}
\caption{Invariant mass of the (anti-)top quark reconstructed at the level of the $W^+W^-b\bar{b}$ final state from the $W$-boson and a $b$-jet with a matching signed flavor tag.}\label{fig:diffxs_ee_mwb}
\end{center}
\end{figure}

Our validation is completed by a comparison of a few selected differential cross sections
in the two subtraction schemes. Fig.~\ref{fig:diffxs_ee_mwb} displays the invariant mass
of the (anti-)top quark reconstructed at the level of the $W^+W^-b\bar{b}$ final state
from the $W$-boson and a $b$-jet with a matching signed flavor tag. The deviation plot
shows excellent statistical compatibility between the two simulations. It also displays
clearly that the pseudo-dipole subtraction technique generates smaller
statistical uncertainties than the standard CS subtraction method.

\subsubsection{$pp \to W^+W^-j_bj_b$}
\begin{figure}[tbp]
\begin{center}
\includegraphics[width=0.625\textwidth]{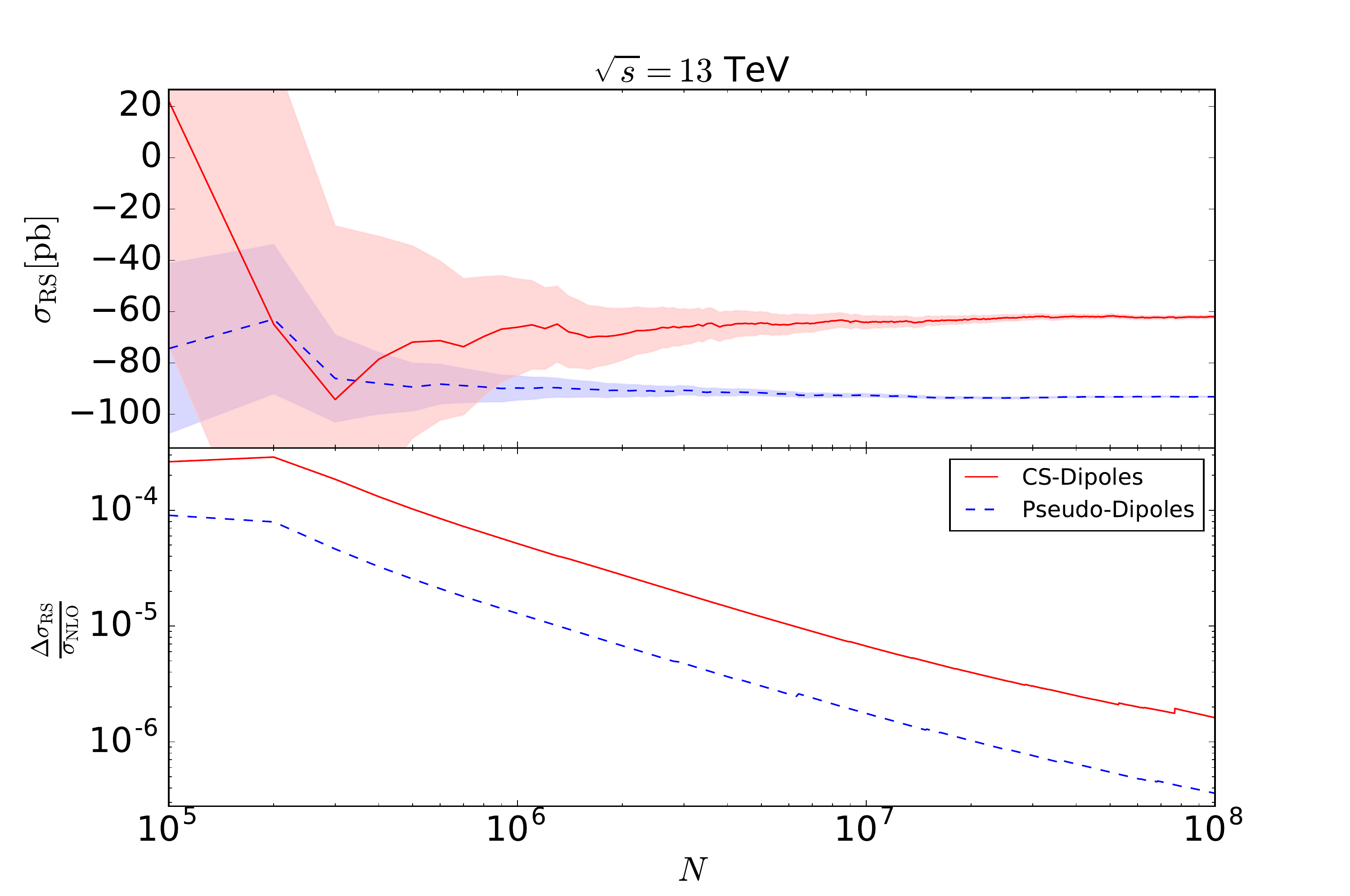}\\
\caption{Evolution of the Monte-Carlo integration results for the subtracted real-emission
  contribution to the total cross section in $pp \to W^+W^-j_bj_b$ at  $\sqrt{s} = 13$~TeV over number of sampled phase-space points $N$. Red solid lines show results from standard CS-dipoles, while
  blue dashed lines correspond to pseudo-dipoles. The colored bands in the upper panels
  and the lines in the lower panels show the one $\sigma$ statistical uncertainty
  of the Monte-Carlo integration.}\label{fig:stat_pp_bwbw}
\end{center}
\end{figure}

As a second application we consider the reaction $pp \to W^+W^-j_bj_b$, where $j_b$ indicates a $b$-tagged jet
with $p_T>25$~GeV. We use the anti-$k_T$ jet algorithm~\cite{Cacciari:2008gp} with $R=0.4$.
The total cross section and the individual contributions to the RS- and the BVI-cross-section for both standard CS- and pseudo-dipoles are given in Tab.~\ref{tab:RSBVI_pp}. Again it can be seen that the convergence of the Monte-Carlo integration of the RS-cross-section for pseudo-dipoles is significantly better than the one for standard CS-dipoles. This can also be observed in Fig.~\ref{fig:stat_pp_bwbw}, which is the analogue of Fig.~\ref{fig:stat_ee_bwbw} for the proton-proton initial state for a center-of-mass energy of $\sqrt{s}=13$~TeV.
\begin{table}[t]
\begin{center}
  \begin{tabular}{|c|c|c|}\hline
    &\multicolumn{2}{c|}{$\sqrt{s}=13$~TeV} \\
    \cline{2-3}
$\sigma\mathrm{[pb]}$ & CS & ID  \\
\hline
RS & $-62.03(59)$ & $-93.21(13)$  \\
BVI & $360.02(39)$ & $391.53(39)$ \\
$\sum$ & $297.99(71)$ & $298.32(41)$ \\\hline
\end{tabular} 
\caption{NLO cross section for $pp \to W^+W^-j_bj_b$ at $\mu_R = \mu_F = m_t$
  and $\sqrt{s}=13$~TeV, computed using standard CS subtraction terms (CS)
  or pseudo-dipoles (ID). The subtracted real-emission contributions (RS) were calculated
  using $10^8$ phase-space points. The Born, virtual corrections and integrated subtraction
  terms (BVI) were calculated using $8.5\cdot10^6$ phase-space-points.
  }\label{tab:RSBVI_pp}
\end{center}
\end{table} 

\begin{figure}[tbp]
\begin{center}
\includegraphics[width=0.49\textwidth]{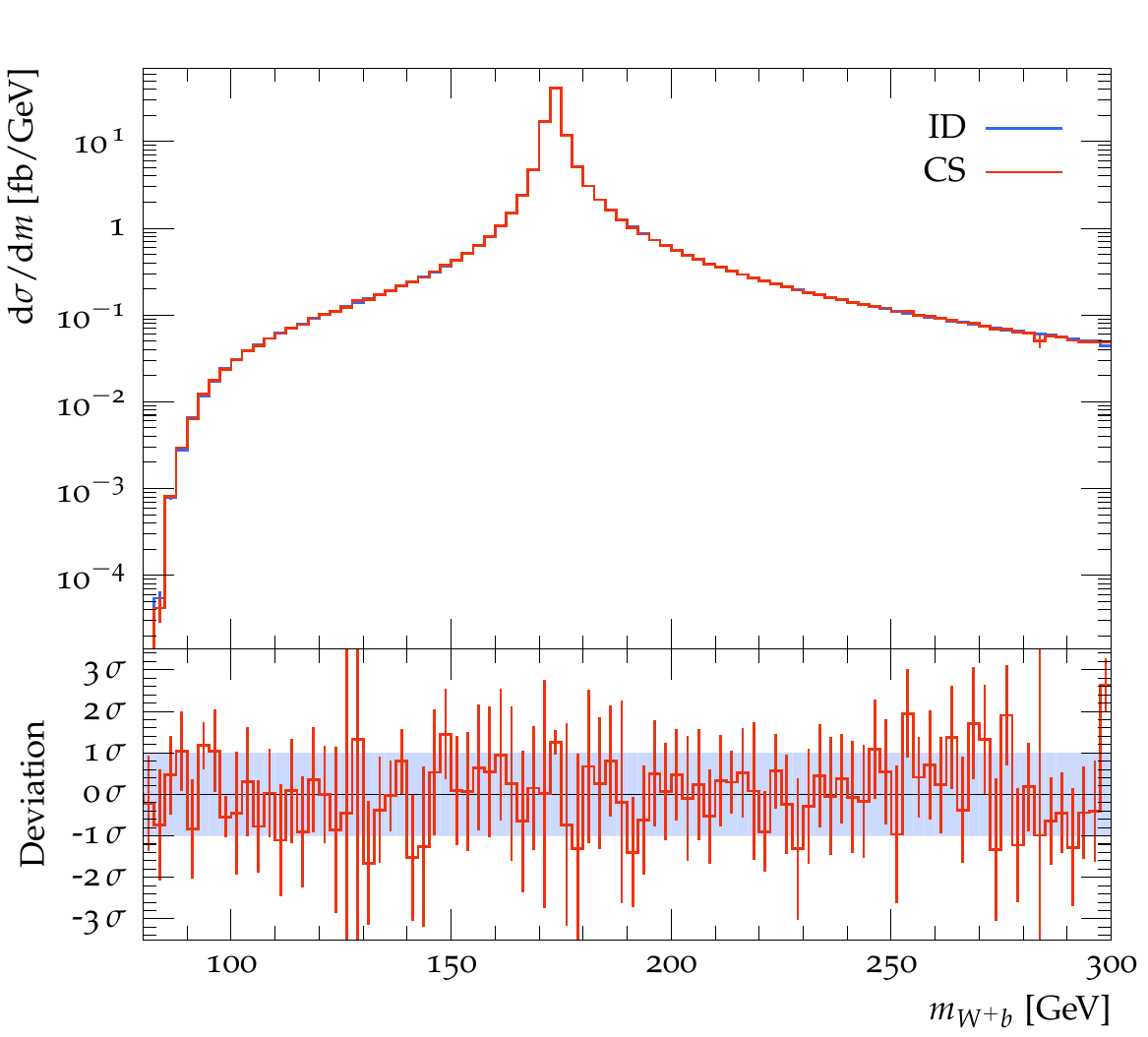}\hfill %
\includegraphics[width=0.49\textwidth]{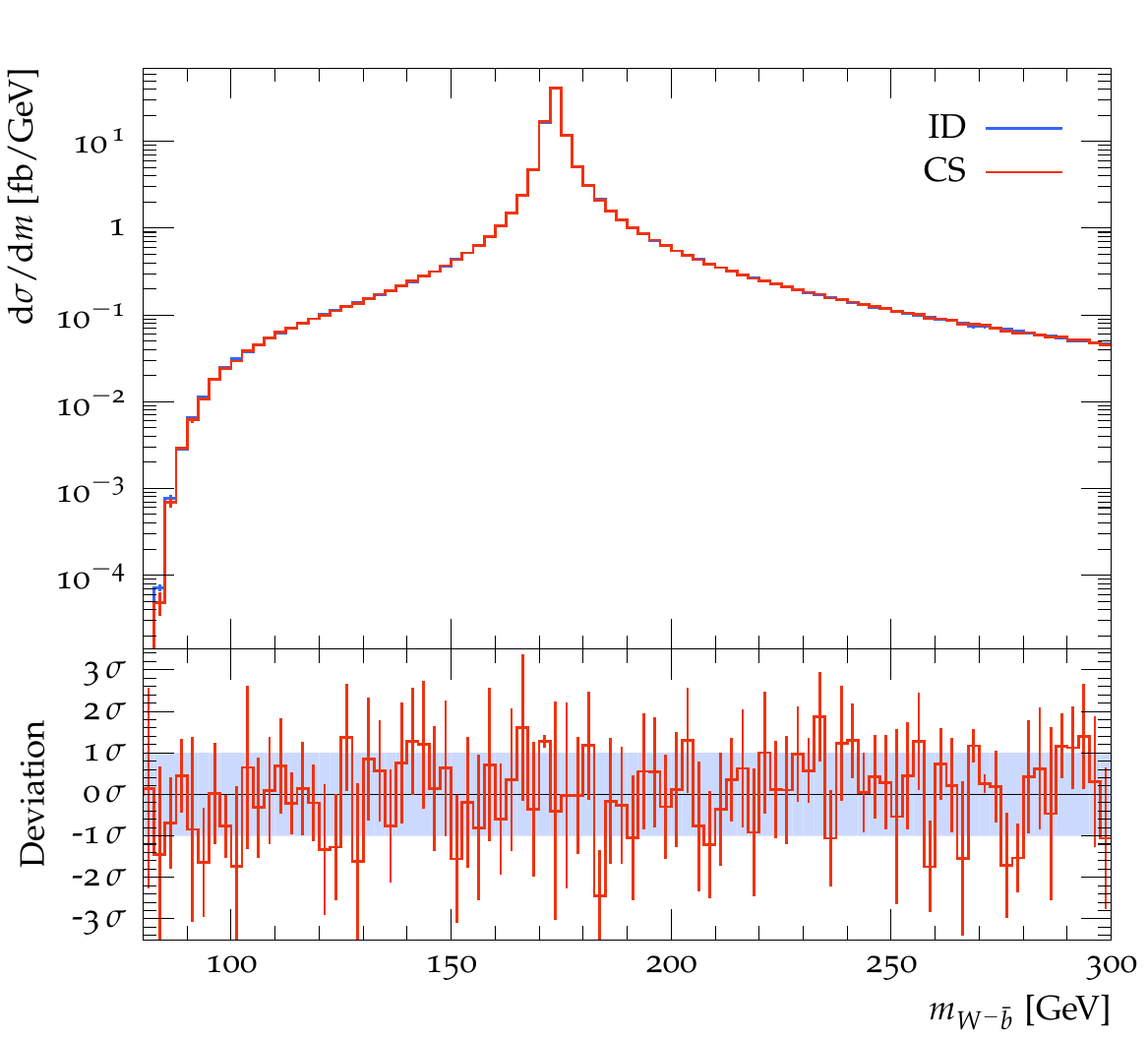}
\caption{Invariant mass of the (anti-)top quark reconstructed at the level of the $W^+W^-j_bj_b$ final state from the $W$-boson and a $b$-jet with a matching flavor tag.}\label{fig:diffxs_pp_mwb}
\end{center}
\end{figure}

Finally, in Fig.~\ref{fig:diffxs_pp_mwb} we confirm the agreement between standard CS-dipoles
and pseudo-dipoles for the differential cross-sections as a function of the combined $W^+j_b$
and $W^-j_{\bar{b}}$ invariant mass. The two subtraction schemes agree within the statistical
uncertainty and pseudo-dipoles exhibit a smaller statistical uncertainty.

\section{Conclusions}
\label{sec:summary}
We have presented a technique that allows to preserve the virtuality
of intermediate propagators in the computation of subtracted real-emission
corrections to processes involving resonances.
We have validated this approach in a simple fixed-order calculation and outlined
how it can be generalized to more complicated processes. Due to the close
correspondence with standard Catani-Seymour dipole subtraction, a matching
to parton showers can be carried out in the MC@NLO or POWHEG methods in the
future, thus paving the way for a precision measurement of processes involving
for example single-top and top-quark pair production.

\appendix

\section{Construction of phase-space trajectories}
\label{sec:momentum_scaling}
In this appendix we describe a generic method to generate phase-space trajectories
approaching the soft or collinear limits of the hard matrix element, as used in
Sec.~\ref{sec:sing_limits}. The technique is based on a suitable scaling of the
Lorentz invariants parametrizing the desired limit. The corresponding kinematical
configuration is determined by combining the appropriate final-state momenta using
the massive dipole kinematics of~\cite{Catani:2002hc} and subsequently constructing
a new real-emission configuration by applying the techniques in~\cite{Hoeche:2009xc}.

\begin{table}
  \begin{center}
  \begin{tabular}{|c|c|c|c|c|c|}\hline
    \multicolumn{3}{|c|}{Initial state} &\multicolumn{3}{c|}{Final state} \\
    \multicolumn{3}{|c|}{$a\to\rm emit$, $b\to\rm spec$} &
    \multicolumn{3}{c|}{$i\to g$, $j\to W$, $k\to b$} \\\cline{2-3}\cline{5-6}
    & soft & collinear & & soft & collinear \\\hline
    $v_{i}'$ & $\lambda v_i$ & $\lambda v_i$ & $y'$ & $\lambda y$ &
    $\displaystyle y\,\frac{1-\lambda \tilde{z}(1-y)}{1-\tilde{z}(1-y)}$ \\
    $x_{i,ab}'$ & $1-\lambda(1-x_{i,ab})$ & $x_{i,ab}$ & $\tilde{z}'$ &
    $\displaystyle\lambda\tilde{z}\frac{1-y}{1-\lambda y}$ &
    $\displaystyle\lambda \tilde{z}(1-y)\left[1-y\,\frac{1-\lambda \tilde{z}(1-y)}{1-\tilde{z}(1-y)}\right]^{-1}$\\\hline
  \end{tabular}
  \end{center}
  \caption{Scaling of Catani-Seymour parameters defined in Eqs.~\eqref{eq:cs_parameters_ff}
    and~\eqref{eq:cs_parameters_ii} used to construct the phase-space trajectories
    in Sec.~\ref{sec:sing_limits}. The scaling parameter is denoted by $\lambda$,
    and the gluon is labeled as particle $i$.\label{tab:scaling}}
\end{table}
We use the notation and definitions of~\cite{Catani:1996vz} for final-state splittings
with final-state spectator
\begin{equation}\label{eq:cs_parameters_ff}
  \tilde{z}_i=\frac{p_ip_k}{(p_i+p_j)p_k}\;,\qquad
  y_{ij,k}=\frac{p_ip_j}{p_ip_j+(p_i+p_j)p_k}\;,
\end{equation}
and for initial-state splitting with initial-state sepctator
\begin{equation}\label{eq:cs_parameters_ii}
  x_{i,ab}=\frac{p_ap_b-p_ip_a-p_ip_b}{p_ap_b}\;,\qquad
  v_i=\frac{p_ip_a}{p_ap_b}\;.
\end{equation}
Table~\ref{tab:scaling} gives the assignment of the momenta to the labels $i$, $j$ and $k$ /
$a$, $b$ and $i$ for the pseudo-dipoles and shows how $y$ and $\tilde{z}$ / $v$ and $x$
are rescaled in order to construct the phase-space trajectories.

The construction of momenta in the final-state splitter case proceeds as follows:
We first combine the three momenta $p_i$, $p_j$ and $p_k$ into intermediate momenta
$\tilde{p}_k$ and $\tilde{p}_{ij}=q-\tilde{p_k}$, where $q=p_i+p_j+p_k$.
\begin{equation}\label{eq:def_ff_clus}
  \begin{split}
    \tilde{p}_k^{\,\mu}=&\;\left(p_k^{\,\mu}-
    \frac{q\cdot p_k}{q^2}\,q^\mu\right)\,
    \sqrt{\frac{\lambda(q^2,m_{ij}^2,m_k^2)}{\lambda(q^2,s_{ij},m_k^2)}}
    +\frac{q^2+m_k^2-m_{ij}^2}{2\,q^2}\,q^\mu\;,
  \end{split}
\end{equation}
where the K{\"a}llen function is given by $\lambda(a,b,c)=(a-b-c)^2-4\,bc$.
Using the scaled parameters $y$ and $\tilde{z}$ from Tab.~\ref{tab:scaling}
we compute the new mass of the pseudoparticle $ij$
as $s_{ij}=y\,(q^2-m_k^2)+(1-y)\,(m_i^2+m_j^2)$, and use
Eq.~\eqref{eq:def_ff_clus} with $\tilde{p}_k\leftrightarrow p_k$
and $m_{ij}\leftrightarrow s_{ij}$ to recalculate $p_k$ and $p_{ij}$.
The new momentum $p_i$ is then constructed as
\begin{align}\label{eq:def_ff_pi_pj}
  p_i^\mu\,=&\;\bar{z}\,\frac{\gamma(q^2,s_{ij},m_k^2)\,p_{ij}^\mu
    -s_{ij}\,p_k^\mu}{\beta(q^2,s_{ij},m_k^2)}
  +\frac{m_i^2+{\rm k}_\perp^2}{\bar{z}}\,
  \frac{p_k^\mu-m_k^2/\gamma(q^2,s_{ij},m_k^2)\,p_{ij}^\mu}{
    \beta(q^2,s_{ij},m_k^2)}+k_\perp^\mu\;,
\end{align}
where $\beta(a,b,c)={\rm sgn}(a-b-c)\sqrt{\lambda(a,b,c)}$ and
$2\,\gamma(a,b,c)=(a-b-c)+\beta(a,b,c)$.
The parameters $\bar{z}$ and ${\rm k}_\perp^2=-k_\perp^2$ 
of this decomposition are given by
\begin{equation}\label{eq:def_ff_zi_kt}
  \begin{split}
    \bar{z}\,=&\;\frac{q^2-s_{ij}-m_k^2}{\beta(q^2,s_{ij},m_k^2)}\,
    \left[\;\tilde{z}\,-\,\frac{m_k^2}{\gamma(q^2,s_{ij},m_k^2)}
      \frac{s_{ij}+m_i^2-m_j^2}{q^2-s_{ij}-m_k^2}\right]\;,\\
        {\rm k}_\perp^2\,=&\;\bar{z}\,(1-\bar{z})\,s_{ij}-
        (1-\bar{z})\, m_i^2-\bar{z}\, m_j^2\;,
  \end{split}
\end{equation}
The transverse momentum is constructed using an azimuthal angle, $\phi_i$
\begin{equation}\label{eq:likt}
  k_\perp^\mu={\rm k}_\perp\left(\cos\phi_i \frac{n_\perp^\mu}{|n_\perp|}
  +\sin\phi_i \frac{l_\perp^{\,\mu}}{|l_\perp|}\right)\;,
\end{equation}
where
\begin{equation}
  n_\perp^\mu=\epsilon^{0\mu}_{\;\;\;\nu\rho}\,
  \tilde{p}_{ij}^{\,\nu}\,\tilde{p}_k^{\,\rho}\;,
  \qquad
  l_\perp^{\,\mu}=\epsilon^\mu_{\;\nu\rho\sigma}\,
  \tilde{p}_{ij}^{\,\nu}\,\tilde{p}_k^{\,\rho}\,n_\perp^\sigma\;.
\end{equation}
In kinematical configurations where $\vec{\tilde{p}}_{ij}=\pm\vec{\tilde{p}}_k$,
$n_\perp$ defined as in Eq.~\eqref{eq:likt} vanishes.
It can then be computed as $n_\perp^\mu=\epsilon^{0\,i\mu}_{\;\;\;\;\;\nu}\,\tilde{p}_{ij}^{\,\nu}$,
where $i$ may be any index that yields a nonzero result.

In the case of initial-state splitters, the construction proceeds as follows:
We first combine the two momenta $p_a$ and $p_i$ into the intermediate momentum
$\tilde{p}_{ai}$.
\begin{equation}\label{eq:def_ii_clus}
  \begin{split}
    \tilde{p}_{ai}^{\,\mu}=&\;\left(p_a^{\,\mu}-
    \frac{m_a^2}{\gamma(s_{ab},m_a^2,m_b^2)}\,p_b^{\,\mu}\right)\,
    \sqrt{\frac{\lambda(q^2,\tilde{m}_{ai}^2,m_b^2)}{
        \lambda(s_{ab},m_a^2,m_b^2)}}+\frac{\tilde{m}_{aij}^2
    }{\gamma(q^2,\tilde{m}_{aij}^2,m_b^2)}\,p_b^{\,\mu}\;,
  \end{split}
\end{equation}
where $q=p_a+p_b-p_i$ and $s_{ab}=(p_a+p_b)^2$. We then apply a
Lorentz transformation to all final-state momenta as
$\tilde{p}^\mu=\Lambda^\mu_{\;\nu} p^\nu$, where
\begin{equation}\label{eq:def_ii_boost}
  \Lambda(\tilde{q},q)^\mu_{\;\nu}=g^\mu_{\;\nu}
  -\frac{2\,(q+\tilde{q})^\mu(q+\tilde{q})_\nu}{(q+\tilde{q})^2}
  +\frac{2\,\tilde{q}^\mu q_\nu}{q^2}\;,
\end{equation}
Using the scaled parameters $v$ and $x$ from Tab.~\ref{tab:scaling}
we compute the new invariant mass squared of the initial state as
$s_{ab}=(q^2-m_i)^2/x_{i,ab}-(m_a^2+m_b^2)(1-x_{i,ab})/x_{i,ab}$
and use the inverse transformation of Eq.~\eqref{eq:def_ii_clus}
to construct the initial-state momentum $p_a$. The new momentum
$p_i$ is then defined as
\begin{align}\label{eq:def_ii_pi_pj}
  \begin{split}
  p_i^{\,\mu}\,=&\;(1-\bar{z}_{ai})\,\frac{
    \gamma(s_{ab},m_a^2,m_b^2)\,p_a^{\,\mu}-m_a^2\,p_b^{\,\mu}}{
    \beta(s_{ab},m_a^2,m_b^2)}\\&\;+
  \frac{m_i^2+{\rm k}_\perp^2}{1-\bar{z}_{ai}}\,\frac{
    p_b^{\,\mu}-m_b^2/\gamma(s_{ab},m_a^2,m_b^2)\,p_a^{\,\mu}}{
    \beta(s_{ab},m_a^2,m_b^2)}-k_\perp^\mu\;,
  \end{split}
\end{align}
with the transverse components constructed according to Eq.~\eqref{eq:likt}.
The parameters $\bar{z}_{aij}$ and ${\rm k}_\perp^2=-k_\perp^2$ 
of this decomposition are given by
\begin{equation}\label{eq:def_ii_zi_kt}
  \begin{split}
    \bar{z}_{ai}\,=&\;\frac{s_{ab}-m_a^2-m_b^2}{\beta(s_{ab},m_a^2,m_b^2)}\,
    \left[\;x_{i,ab}+v_i\,-\,\frac{m_b^2}{\gamma(s_{ab},m_a^2,m_b^2)}
      \frac{s_{ai}+m_a^2-m_i^2}{s_{ab}-m_a^2-m_b^2}\right]\;,\\
    {\rm k}_\perp^2\,=&\;\bar{z}_{ai}\,(1-\bar{z}_{ai})\,m_a^2
    -(1-\bar{z}_{ai})\,s_{ai}-\bar{z}_{ai}\,m_i^2\;,
  \end{split}
\end{equation}
Finally we boost all remaining final-state particles according to
Eq.~\eqref{eq:def_ii_boost} with $q\leftrightarrow\tilde{q}$.

\bibliographystyle{JHEP}
\bibliography{bibliography}

\end{document}